\documentclass[prl,reprint,superscriptaddress,nofootinbib]{revtex4-2}

\usepackage{amsmath,amssymb,bm,graphicx}
\usepackage[colorlinks=true,citecolor=blue,linkcolor=blue,urlcolor=blue]{hyperref}

\newcommand{\mathbfcal}[1]{\bm{\mathcal{#1}}}
\newcommand{\intq}{\int_{\bm q}}
\newcommand{\intqp}{\int_{\bm q'}}

\usepackage{xcolor}
\definecolor{PRLblue}{RGB}{0,55,120}
\newcommand{\PRLsection}[1]{%
  \par\vspace{0.em}%
\noindent{\color{PRLblue}\itshape #1.—}\ %
}

\begin{document}

\title{Nonlocal Orbital-Angular-Momentum Dichroism of Vortex Light in Strained Crystals}

\author{Habib Rostami}
\email{hr745@bath.ac.uk}
\affiliation{Department of Physics, University of Bath, Claverton Down, Bath BA2 7AY, United Kingdom}

\date{\today}

\begin{abstract}
Vortex light carries orbital angular momentum (OAM) in its transverse phase. 
We show that the transverse phase cannot generate OAM-odd absorption in any continuously translation-invariant system, even
when optical nonlocality and multipole light--matter coupling are included. 
The combination of broken continuous translation symmetry and finite-range optical nonlocality permits OAM dichroism through transverse-momentum mixing; in nonuniformly strained Dirac materials, the dichroism factorizes into a nonlocal elasto-optic tensor and the overlap of the strain texture with the vortex transverse phase current.
\end{abstract}

\maketitle

\PRLsection{Introduction}
Optical vortices carry orbital angular momentum (OAM) through the transverse
phase \(e^{i\ell\phi}\), with \(L_z=\ell\hbar\) per photon
~\cite{Allen1992,YaoPadgett2011,Shen2019}. Unlike spin angular momentum (SAM),
which is encoded locally in circular polarization, OAM is encoded in transverse
phase coherence and can access higher multipole light--matter couplings
~\cite{QuinteiroRMP2022}. This transverse structure enables vortex light to
drive higher-order selection rules for OAM transfer and chiroptical effects in
atomic, molecular, and nanostructured systems
~\cite{Andersen2006,Walker2012,Schmiegelow2016,DeNinno2020,Begin2025,
Babiker2002,QuinteiroBerakdar2009,Loffler2011,Brullot2016,
ForbesAndrews2021,ForbesJones2021}.
OAM-sensitive condensed-matter and nanophotonic responses are explored in semiconductor
transitions
~\cite{QuinteiroRMP2022,QuinteiroTamborenea2009,QuinteiroTamborenea2010},
photocurrent generation~\cite{Ji2020}, excitonic OAM imprinting
~\cite{Simbulan2021,Kesarwani2022,Hafezi2022,GomezSanchez2024}, single-photon emission
~\cite{Zhang2023OAMEmitter}, magnetic helicoidal dichroism
~\cite{Fanciulli2022}, superconducting-vortex control~\cite{Yeh2024},
quantum-Hall OAM pumping~\cite{Session2025}, and on-chip OAM
photodetection~\cite{Dai2024OnChip}.

Conventional SAM dichroism provides the local \mbox{\(q=0\)} benchmark where circular polarization
couples to the optical Hall conductivity and probes Berry curvature, valley selectivity, orbital magnetization, topological band
structure, and Kerr-active hidden order ~\cite{XiaoRMP2010,XiaoTMD2012,Cao2012,Mak2012,Zeng2012,XuNP2014,Rostami2019,Cappelluti2025HiddenOrderKerr}.
For circular polarization vector \(\bm e_s=(\bm e_x+is\bm e_y)/\sqrt2\) with spin $s=+/-$ for right/left-handedness, the SAM-projected conductivity is
\mbox{\(\sigma_s=e^*_{s,i}\sigma_{ij}e_{s,j}=\sigma_{xx}+is\sigma_{xy}\)}. Thus the SAM-odd absorption contrast
\(\Delta \sigma^{\rm SAM}_s\equiv{\rm Re}(\sigma_{s}-\sigma_{-s})\propto s {\rm Im}\,\sigma_{xy}\),
isolating a local antisymmetric Hall conductivity.
A local conductivity samples only \(E_i^*(\bm r)E_j(\bm r)\), which is
even under \(\ell\to-\ell\). Therefore, distinguishing an OAM dichroism requires a nonlocal response \cite{AgranovichGinzburg1984,Urru2025,KatoYokoshi2026}. 
However, as we show below, nonlocality is insufficient: transverse-momentum conservation imposed by translational invariance removes the OAM sensitivity of absorption.
 \begin{figure}
    \centering
\includegraphics[width=1\linewidth]{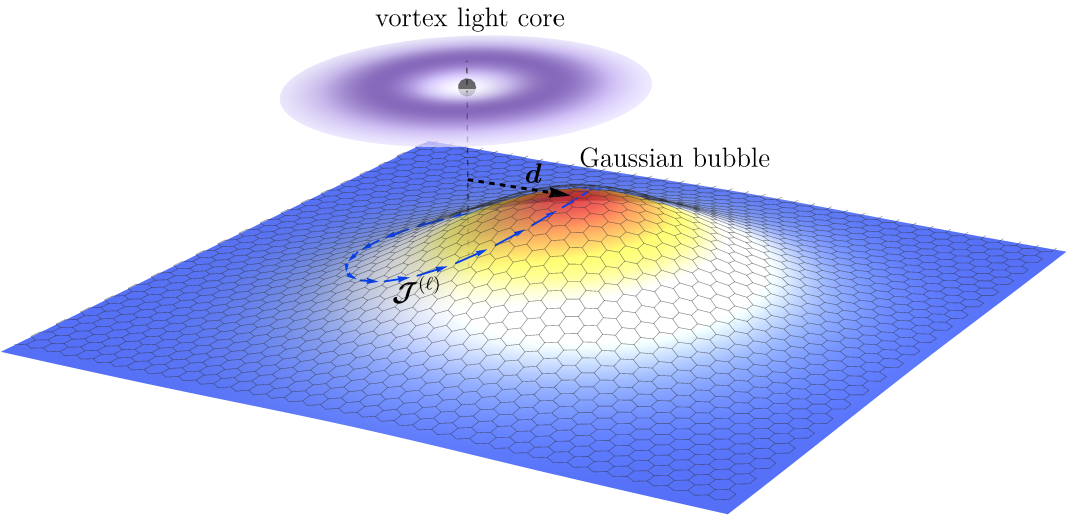}
\caption{
Schematic of an optical vortex incident on a nonuniformly strained hexagonal
2D material. The surface represents a Gaussian strain bubble, with
colour indicating the deformation amplitude. The vortex core is displaced from
the bubble center by the in-plane vector \(\bm d\). The circular arrows denote
the vortex transverse phase current \(\mathbfcal J^{(\ell)}\). This relative
displacement breaks the common circular symmetry of the optical and strain
profiles, enabling OAM-sensitive coupling through optical nonlocality.
}
    \label{fig1}
\end{figure}
\par
In this Letter, we show that OAM absorption dichroism (OAMD) of scalar vortex light with
fixed polarization is forbidden in both local optical response and translationally invariant nonlocal response. 
It requires broken continuous translational symmetry, supplied by a nonuniform spatial texture such as strain, heterostrain, defects, boundaries, relaxed moiré modulation, or crystalline Umklapp scattering. As an example, we realize this condition in a nonuniformly strained two-dimensional (2D) Dirac
system where the  spatial dispersion (nonlocality) supplies the average optical momentum dependence,
while the nonuniform strain supplies transverse momentum
transfer. Hexagonal symmetry enters through a valley-odd pseudo-gauge coupling, selecting
the OAM-odd tensor channels and requiring valley inequivalence.
The result separates optical geometry from material response. For a Laguerre--Gaussian (LG) vortex beam, the OAM dichroism takes the factorized
form
\begin{equation}
    \Delta \sigma^{\rm OAM}_{\ell, s}
    \equiv
    {\rm Re}
    \left[
     {\cal B}^{(\ell)}_{ab}
        \Lambda^{(s)}_{ab}(\omega)
    \right],
    \label{eq:intro_OAM_overlap}
\end{equation}
where repeated Cartesian indices \(a,b=x,y\) are summed. The OAM-projected {\em geometric overlap tensor} (GOT) 
\begin{equation}
    {\cal B}^{(\ell)}_{ab}
    =
    \int_{\bm r}
    {\cal A}_a(\bm r)
    {\cal J}^{(\ell)}_b(\bm r),
    \label{eq:intro_overlap_tensor}
    \vspace{-2mm}
\end{equation}
with \({\cal A}_a(\bm r)\) the strain-induced vector field and
\({\cal J}^{(\ell)}_b(\bm r)\) the vortex  {\em transverse phase current} (TPC),
\begin{equation}
    \mathbfcal{J}^{(\ell)}(\bm r)
    =
    {\rm Im}
    \left[
        u_\ell^*(\bm r)\bm\nabla u_\ell(\bm r)
    \right],
    \label{eq:intro_OAM_current}
\end{equation}
where \(u_\ell(\bm r)\) is the LG profile carrying OAM charge \(\ell\). The material factor $\Lambda^{(s)}_{ab}(\omega)$ is the SAM-projected {\em nonlocal elasto-optic} (NLEO) response  that controls the part of the optical conductivity that is linear in the strain field and in optical spatial dispersion. 
Equations~\eqref{eq:intro_OAM_overlap}–\eqref{eq:intro_OAM_current} form the central results which factorize the
OAMD into the SAM-projected NLEO tensor,
\(\Lambda^{(s)}_{ab}\), and the OAM-projected GOT, \({\cal B}^{(\ell)}_{ab}\).
Their contraction selects the transition channels that are sensitive to SAM, OAM,
the elastic texture, and the electronic anisotropy. We first establish the \emph{no-go limits} of OAM dichroism and then compute the response tensors for tilted massive Dirac electrons in a 2D crystal with a Gaussian  bubble, as shown in
Fig.~\ref{fig1}.

\PRLsection{Vortex-light OAM dichroism}
We start from the nonlocal optical response
\begin{equation}
    j_i(\bm r,\omega)
    =
    \int_{\bm r'} \,
    \sigma_{ij}(\bm r,\bm r',\omega)E_j(\bm r',\omega),
    \label{eq:nonlocal_response}
\end{equation}
where $j_i(\bm r,\omega)$ is the charge current generated in response to a vortex light field $E_j(\bm r',\omega)$ and \(\sigma_{ij}(\bm r,\bm r',\omega)\) is the fully nonlocal dynamical conductivity kernel of a 2D material. The absorbed power is
\begin{equation}
    P_{\rm abs}
    =
    \frac{1}{2}{\rm Re}
    \int_{\bm r} \,E_i^*(\bm r,\omega)j_i(\bm r,\omega).
    \label{eq:power}
\end{equation}
We consider a single-mode vortex beam with OAM index \(\ell\) and SAM index
\(s\), incident normally on the 2D material. Since the material is atomically thin and the height variation is negligible on the optical wavelength scale, we evaluate the beam profile at the waist plane with an electric
field 
$ \bm E_{\ell, s}(\bm r,t)
    =
    {\rm Re}
    \left[
        {\cal E}_0
        u_\ell(\bm r)
        \bm e_s
        e^{-i\omega t}
    \right]
$~\cite{Allen1992},
where \(\bm r=(x,y)\) is the transverse coordinate. We use the lowest-radial-index LG mode
\begin{equation}
    u_\ell(\bm r)
    =
    N_\ell
    \left(\frac{r}{w_0}\right)^{|\ell|}
    e^{-r^2/w_0^2}
    e^{i\ell\phi},
    \label{eq:LG_envelope}
\end{equation}
with beam waist \(w_0\), polar angle \(\phi\), and normalization
\(N_\ell=[2^{|\ell|+1}/(\pi|\ell|!w_0^2)]^{1/2}\), such that
\(\int_{\bm r} |u_\ell(\bm r)|^2=1\). The 2D Fourier transform of the mode profile is
$ \tilde u_\ell(\bm q)
    =
    \int_{\bm r}
    u_\ell(\bm r)e^{-i\bm q\cdot\bm r}
    =
    f_\ell(q)e^{i\ell\phi_{\bm q}}$,
up to an irrelevant global phase~\cite{LGGlobalPhase}. For the convention in
Eq.~\eqref{eq:LG_envelope},
\begin{equation}
    f_{\ell}(q)
    =
    \tilde N_\ell
    (w_0q)^{|\ell|}
    e^{-w_0^2q^2/4},
    \label{eq:LG_momentum}
\end{equation}
where
\(\tilde N_\ell=[\pi w_0^2/(2^{|\ell|-1}|\ell|!)]^{1/2}\), and
\(\phi_{\bm q}\) is the polar angle of \(\bm q\).
The measurable quantity is the bilinear overlap between the optical mode and the nonlocal response. For a normalized mode \((\ell,s)\), we define mode-projected optical response in position and momentum space: 
\begin{align}
    \Sigma_{\ell, s}(\omega)
    &=
    \int_{\bm r}\int_{\bm r'} \,
    u^\ast_{\ell}(\bm r)
    e^*_{s,i}
    \sigma_{ij}(\bm r,\bm r',\omega)
    e_{s,j}
    u_{\ell}(\bm r'),\nonumber\\
   & =
    \int_{\bm q}\int_{\bm q'} \,
    \tilde u^\ast_{\ell}(\bm q)
    e^*_{s,i}
    \sigma_{ij}(\bm q,\bm q',\omega)
    e_{s,j}
   \tilde u_{\ell}(\bm q').
    \label{eq:projected_sigma}
\end{align}
The absorption is governed by \(\frac{1}{2}{\rm Re}\,\Sigma_{\ell s}(\omega)\) and thus we define the OAMD contrast as
\begin{equation}
    \Delta \sigma^{\rm OAM}_{\ell,s}
    =
    \frac{1}{2}
    {\rm Re}
    \left[
        \Sigma_{+\ell,s}(\omega)
        -
        \Sigma_{-\ell,s}(\omega)
    \right].
\end{equation}
Analogous SAM-odd and total-angular-momentum-odd contrasts can be defined, but here
we focus on the OAM-odd response.

\PRLsection{No-go limits}
We now state the basic no-go constraints. For a scalar paraxial beam, Eq.~\eqref{eq:projected_sigma} shows that the sign of the OAM
charge enters only through the relative phase
\(e^{i\ell(\phi'-\phi)}\) or \(e^{i\ell(\phi_{\bm q'}-\phi_{\bm q})}\). An OAM-odd signal therefore requires
the response kernel to compare different transverse momentum components in a
way that does not make this phase even under \(\ell\!\to\!-\!\,\ell\).
It is useful to introduce Wigner variables,
\(\bm R\!=\!(\bm r\!+\!\bm r')/2\), \(\bm\rho\!=\!\bm r\!-\!\bm r'\),
\(\bm Q\!=\!(\bm q\!+\!\bm q')/2\), and \(\bm k\!=\!\bm q\!-\!\bm q'\), so that 
\begin{align}
    \sigma_{ij}(\bm q,\bm q',\omega)
    =\int_{\bm R}\int_{\bm \rho}
    e^{i\bm k\cdot\bm R+i\bm Q\cdot\bm \rho}
    \sigma_{ij}(\bm R,\bm \rho,\omega).
\label{eq:sigma_Wigner_transform}
\end{align}
Here \(\bm k\) is the transverse momentum transferred by the material texture,
whereas \(\bm Q\) probes the relative-coordinate nonlocality of the response. In the following we use \(\sigma\) for both the full conductivity kernel and its reduced forms in special
limits. The arguments specify the object; the physical dimensions may differ
because the accompanying delta functions carry dimensions.

Case 1: {\em local response.}
Spatial inhomogeneity alone is insufficient for OAMD when the response is local
in the relative coordinate:
\begin{equation}
    \sigma_{ij}(\bm R,\bm\rho,\omega)
    =
    \sigma_{ij}(\bm R,\omega)\delta(\bm\rho),
\end{equation}
which gives the mode-projected response
$\Sigma_{\ell,s}(\omega)
    =
    \int_{\bm r}
    |u_\ell(\bm r)|^2
    e^*_{s,i}\sigma_{ij}(\bm r,\omega)e_{s,j}$.
Thus the response samples only the local intensity, so the vortex phase cancels
pointwise and \(\Sigma_{\ell,s}=\Sigma_{-\ell,s}\). The homogeneous
electric-dipole limit follows by dropping the \(\bm R\)-dependence,
\(\sigma_{ij}(\bm R,\omega)\to\sigma_{ij}(\omega)\), equivalently
\(\sigma_{ij}(\bm q,\bm q',\omega)
=(2\pi)^2\delta(\bm q-\bm q')\sigma_{ij}(\omega)\). Hence neither a homogeneous
local response nor a slowly varying local inhomogeneous texture can resolve vortex light
handedness.

Case 2: \emph{homogeneous nonlocal response.}
For any nonlocal but
translationally invariant kernel,
\begin{equation}
    \sigma_{ij}(\bm R,\bm\rho,\omega)
    =
    \sigma_{ij}(\bm\rho,\omega),
    \label{eq:homogeneous_nonlocal_kernel}
\end{equation}
the center coordinate drops out. Equation~\eqref{eq:sigma_Wigner_transform}
then gives transverse momentum conservation, $    \sigma_{ij}(\bm q,\bm q',\omega)
    = (2\pi)^2\delta(\bm q-\bm q')\, \sigma_{ij}(\bm q,\omega)$. 
Thus transverse momentum is conserved, \(\bm q'=\bm q\), and the vortex phase
factor reduces to \(e^{i\ell(\phi_{\bm q'}-\phi_{\bm q})}=1\). Such a response can produce optical activity, photon drag, and
electric-quadrupole/magnetic-dipole corrections, including second-order optical
response in graphene
~\cite{AgranovichGinzburg1984,WangTokmanBelyanin2016,ChengVermeulenSipe2017,RostamiKatsnelsonPolini2017},
but remains OAM-even because translational symmetry forbids mixing distinct
transverse components of the vortex beam.

Case 3: {\em inhomogeneous infinite-range response.}
An idealized nonlocal limit is
\begin{equation}
      \sigma_{ij}(\bm R,\bm\rho,\omega)=\sigma_{ij}(\bm R,\omega),  
\end{equation}
which is inhomogeneous but independent of the relative coordinate. The kernel
therefore couples equally to all separations in \(\bm\rho\), i.e. it is
infinite range. Equation~\eqref{eq:sigma_Wigner_transform} gives $   \sigma_{ij}(\bm q,\bm q',\omega)
    = (2\pi)^2\delta(\bm Q)\sigma_{ij}(\bm k,\omega)$. 
Thus \(\bm Q=0\), or \(\bm q'=-\bm q\), and the vortex phase becomes
\(e^{i\ell(\phi_{\bm q'}-\phi_{\bm q})}=(-1)^\ell\), which is even under
\(\ell\to-\ell\). 
Interaction with a uniform bosonic mode may generate such an infinite-range kernel~\cite{Rostami2018}, but the response remains OAM-even.


Case 4: {\em inhomogeneous nonlocal response.}
The allowed class is a kernel with genuine dependence on both Wigner
variables, \(\sigma_{ij}(\bm R,\bm\rho,\omega)\): \(\bm R\)-dependence breaks
continuous translational symmetry and supplies transverse momentum transfer
\(\bm k=\bm q-\bm q'\), while \(\bm\rho\)-dependence gives optical
nonlocality, encoded by \(\bm Q=(\bm q+\bm q')/2\). Together they allow the
response to compare distinct transverse components of a scalar vortex beam
without imposing either \(\bm q'=\bm q\) or \(\bm q'=-\bm q\).
Examples include separable kernels from finite-size or long-wavelength
modulations,
\begin{equation}
    \sigma_{ij}(\bm R,\bm\rho,\omega)
    =
    \sigma_{ija}(\bm\rho,\omega)w_a(\bm R),
\end{equation}
which give
\(\sigma_{ij}(\bm q,\bm q',\omega)
=\sigma_{ija}(\bm Q,\omega)w_a(\bm k)\), and localized nonlocal scatterers,
\(\sigma_{ij}=\sigma_{ij}(\bm\rho,\omega)\delta(\bm R-\bm R_\ast)\), for which
\(\sigma_{ij}(\bm q,\bm q',\omega)
=e^{i\bm k\cdot\bm R_\ast}\sigma_{ij}(\bm Q,\omega)\). Neither imposes the momentum constraints above, so an OAM-odd signal is allowed, though not
guaranteed.
Finally, crystalline Umklapp can also supply \(\bm q-\bm q'=\bm G\), but
atomic reciprocal vectors lie outside the transverse-momentum support of a
vortex beam because \(|\tilde u_\ell(q)|\) is exponentially suppressed for
\(q\gtrsim1/w_0\), see Eq.~\eqref{eq:LG_momentum}. Efficient OAMD thus
requires a mesoscopic texture comparable to the beam waist.

\begin{figure}
    \centering    \includegraphics[width=1\linewidth]{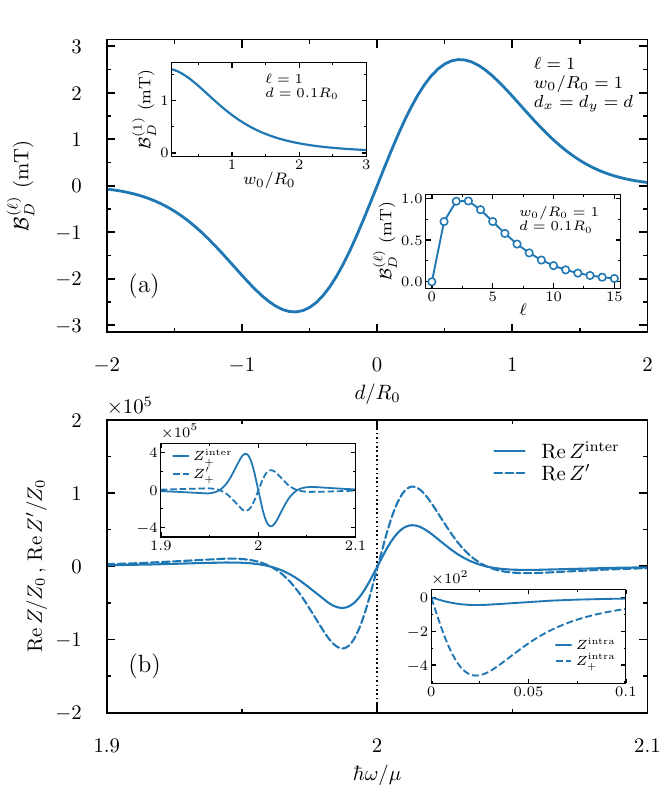}
\caption{
OAM-projected geometric overlap integral ${\cal B}^{(\ell)}_{ab}$ and
SAM-projected material tensor \(\Lambda^{(s)}_{ab}\) for strain-assisted OAM
dichroism.
(a) Geometric overlap for a displaced Gaussian bubble, shown through
\mbox{\({\cal B}^{(\ell)}_D={\cal B}^{(\ell)}_{xx}
-{\cal B}^{(\ell)}_{yy}\)}. For \mbox{\(d_x=d_y=d\)}, the symmetric
off-diagonal component
\mbox{\({\cal B}^{(\ell)}_S={\cal B}^{(\ell)}_{xy}
+{\cal B}^{(\ell)}_{yx}\)} is numerically identical to
\mbox{\({\cal B}^{(\ell)}_D\)}. The main panel shows
\mbox{\({\cal B}^{(1)}_D\)} versus \mbox{\(d/R_0\)} for
\mbox{\(\ell=1\)} and \mbox{\(w_0=R_0\)}. Insets show its dependence on
\mbox{\(w_0/R_0\)} at \mbox{\(\ell=1\)},
\mbox{\(d_x=d_y=0.1R_0\)}, and on \mbox{\(\ell\)} at
\mbox{\(w_0=R_0\)}, \mbox{\(d_x=d_y=0.1R_0\)}. We use a Gr\"uneisen
parameter \mbox{\(\beta\approx2\)} and the universal bubble aspect ratio
\mbox{\(h_0/R_0\approx0.1\)}~\cite{Khestanova2016UniversalBubbles}. We set
\mbox{\(R_0=1\,\mu{\rm m}\)}.
(b) Material response near the interband resonance
\mbox{\(\hbar\omega\approx2\mu\)}, normalized by
\mbox{\(Z_0= e^3/(2\pi\mu^2)\)}. The main panel shows the two-valley
valley-odd responses \mbox{\({\rm Re}\,Z_{\rm inter}\)} and
\mbox{\({\rm Re}\,Z'\)} for \mbox{\(\mu_+=\mu_-=\mu\)},
\mbox{\(m_S/\mu=0.20\)}, \mbox{\(m_H/\mu=0.08\)}, and
\mbox{\(\gamma/\mu=0.04\)}, with \mbox{\(m_\pm=m_S\pm m_H\)}. The
upper-left inset shows the corresponding \(\tau=+\) valley interband contributions
\mbox{\({\rm Re}\,Z^{\rm inter}_{+}/Z_0\)} and
\mbox{\({\rm Re}\,Z'_{+}/Z_0\)} on the same frequency window. The
lower-right inset shows the low-frequency intraband response, comparing the
two-valley contribution \mbox{\({\rm Re}\,Z^{\rm intra}/Z_0\)} with its
\(\tau=+\) valley component \mbox{\({\rm Re}\,Z^{\rm intra}_{+}/Z_0\)}.
}\label{fig2}
\vspace{-3mm}
\end{figure}

Table~\ref{tab:no_go_kernels} summarizes the no-go cases. OAMD requires
a spatially inhomogeneous nonlocal response; it is therefore a genuinely
nonlocal, texture-assisted effect.
This requirement is consistent with existing OAM-sensitive solid-state
experiments. Magnetic helicoidal dichroism vanishes for homogeneous magnetic
structures and appears only when a magnetic vortex supplies the texture needed
to mix OAM channels~\cite{Fanciulli2022}. Trapped-exciton emitters require
strain or moir\'e confinement, which makes the exciton center-of-mass motion
carry valley-entangled OAM~\cite{Zhang2023OAMEmitter}. On-chip OAM
photodetectors similarly use a patterned spin-Hall plasmonic coupler to convert
vortex phase into spatially separated hot spots before local electrical readout
~\cite{Dai2024OnChip}. Below we realize the intrinsic material counterpart of
these engineered routes using weak nonuniform strain in a  2D
tilted massive Dirac model. Electrostatic confinement provides an analogous
route by supplying finite-\(\bm k\) components.
\begin{table}[t]
\caption{
Classification of response kernels in Wigner coordinates
\(\bm R=(\bm r+\bm r')/2\) and \(\bm\rho=\bm r-\bm r'\).
OAMD requires broken continuous translational symmetry and finite-range
optical nonlocality; infinite-range response remains OAM-even.
}
\label{tab:no_go_kernels}
\begin{ruledtabular}
\setlength{\tabcolsep}{3pt}
\begin{tabular}{@{}lllll@{}}
Class & Kernel & Inhom. & Nonlocal & OAMD \\
\hline
Local
& \(\sigma(\omega)\delta(\bm\rho)\)
& No & No & No \\
Homo.
& \(\sigma(\bm\rho,\omega)\)
& No & Yes & No \\
Texture
& \(\sigma(\bm R,\omega)\delta(\bm\rho)\)
& Yes & No & No \\
Global
& \(\sigma(\bm R,\omega)\)
& Yes & $\infty$-range & No \\
Separable
& \(\sigma(\bm\rho,\omega)w(\bm R)\)
& Yes & Yes & Possible \\
Localized
& \(\sigma(\bm\rho,\omega)\delta(\bm R-\bm R_\ast)\)
& Yes & Yes & Possible \\
Generic
& \(\sigma(\bm R,\bm\rho,\omega)\)
& Yes & Yes & Possible \\
\end{tabular}
\end{ruledtabular}
\end{table}
\PRLsection{Strain-assisted OAM dichroism}
In 2D materials with hexagonal symmetry, strain generates a
valley-odd pseudo-gauge field \(\mathbfcal A(\bm R)\), which preserves
time-reversal symmetry and is not screened by conduction electrons. Such
pseudo-gauge fields have been probed experimentally through pseudo-Landau levels
and acoustically induced synthetic Hall voltages in graphene
~\cite{Levy2010PseudoLandau,Jiang2017VisualizingPMF,Nigge2019RoomTemperaturePLL,Zhao2022SyntheticHall}.
Here, \(\mathbfcal A(\bm R)\) supplies the transverse momentum transfer required
for OAMD. To leading order in a weak, long-wavelength strain texture,
the nonlocal conductivity takes the Wigner form
\begin{equation}
    \sigma_{ij}(\bm R,\bm\rho,\omega)
    \approx
    \sigma^{(0)}_{ij}(\bm\rho,\omega)
    +
    \sigma^{(1)}_{ija}(\bm\rho,\omega){\cal A}_a(\bm R).
    \label{eq:Wigner_sigma}
\end{equation}
where \(\sigma^{(0)}_{ij}\) is the     homogeneous nonlocal conductivity of the translationally
invariant system, and \(\sigma^{(1)}_{ija}\) is the NLEO response
to the strain-induced pseudo-gauge field. By the no-go result of
Case~2, the first term in Eq.~\eqref{eq:Wigner_sigma} is OAM-even. We therefore
retain the strain-induced term, which in momentum space reads
$\delta\sigma_{ij}(\bm Q,\bm k,\omega)
    \approx
    \sigma^{(1)}_{ija}(\bm Q,\omega){\cal A}_a(\bm k)$. 
This gives the OAM-sensitive part of the mode-projected conductivity,
\begin{equation}
  \hspace{-3mm}  \delta\Sigma_{\ell, s}(\omega)
    \!\approx\!
    \intq\intqp
    \tilde u^\ast_\ell (\bm q) \tilde u_\ell(\bm q')
    {\cal A}_a(\bm k)
    e^*_{s,i}\sigma^{(1)}_{ija}(\bm Q,\omega)e_{s,j}.
    \label{eq:strain_projected_sigma}
\end{equation}
Since the local limit, \(\sigma^{(1)}_{ija}(\bm Q\to0,\omega)\), is
OAM-even, the leading OAM-sensitive contribution arises from the first
spatial-dispersion correction,
\begin{equation}
    \delta\sigma^{(1)}_{ija}(\bm Q,\omega)
    =
    \zeta_{ijab}(\omega)Q_b
    +
    {\cal O}(Q^2).
    \label{eq:gamma_expansion_main}
\end{equation}
Here \(\zeta_{ijab}(\omega)\) is the leading NLEO  tensor, i.e. the electric-quadrupole/magnetic-dipole optical response. Substituting Eq.~\eqref{eq:gamma_expansion_main} into Eq.~\eqref{eq:strain_projected_sigma} and transforming to real space gives~\cite{sm}
\begin{equation}
\label{eq:dSigma}
\hspace{-2mm}\delta\Sigma_{\ell,s}(\omega)
=
{\cal B}^{(\ell)}_{ab}\Lambda^{(s)}_{ab}(\omega),
\quad
\Lambda^{(s)}_{ab}(\omega)
=
e^*_{s,i}\zeta_{ijab}(\omega)e_{s,j}.
\end{equation}
This yields the OAM-odd contrast in Eq.~\eqref{eq:intro_OAM_overlap}, factorized into the vortex--strain overlap \({\cal B}^{(\ell)}_{ab}\) defined in Eq.~\eqref{eq:intro_overlap_tensor} and the SAM-projected elasto-optic nonlocal response \(\Lambda^{(s)}_{ab}\).
\PRLsection{OAM-projected geometric overlap tensor}
Bubble-shaped deformations provide smooth, mesoscopic strain textures in
2D materials
~\cite{Levy2010PseudoLandau,CastellanosGomez2013LocalStrain,
Khestanova2016UniversalBubbles,Blundo2021StrainTuning}.
In van der Waals heterostructures, such bubbles span tens of nanometers to
microns and have an approximately universal height-to-radius ratio
~\cite{Khestanova2016UniversalBubbles}. For a hexagonal Dirac material, strain enters the low-energy Hamiltonian as a
valley-odd pseudo-gauge field
$
\mathbfcal A=A_0(\varepsilon_{xx}-\varepsilon_{yy},-2\varepsilon_{xy}),
$
where \(A_0=\beta\phi_B/(2\pi a_0)\), \(\beta\) is the Gr\"uneisen parameter,
\(\phi_B=\pi\hbar/e\) is the magnetic flux quantum, and \(a_0\) is the
nearest-neighbor bond length
~\cite{Rostami2015TheoryStrainTMD,Rostami2018PiezoelectricityValleyChern}. The strain tensor is
\(\varepsilon_{ij}=(\partial_i\xi_j+\partial_j\xi_i+\partial_i h\,\partial_j h)/2\),
with \(\bm\xi\) and \(h\) the in-plane and out-of-plane displacements. We model the relaxed Gaussian bubble as displaced from the vortex core by a vector \(\bm d\), such that
\(\mathbfcal{A}(\bm R)=\mathbfcal{A}^{G}(\bm R-\bm d)\).
The field \(\mathbfcal{A}^{G}\) is generated by the rotationally symmetric Gaussian out-of-plane deformation
\(h(R)=h_0\exp[-R^2/(2R_0^2)]\) with corresponding gauge field 
\(\mathbfcal{A}^{G}(\bm R)={\cal A}^{G}(R)(\cos2\phi,-\sin2\phi)\),
where, neglecting the Poisson-ratio correction, we have~\cite{Rostami2018PiezoelectricityValleyChern}
\begin{equation}
{\cal A}^{G}(R)
=
R_0{\cal B}_0
\frac{
[1+(R/R_0)^2]e^{-R^2/R_0^2}-1
}{
(R/R_0)^2
},
\label{eq:gaussian_bubble_A}
\end{equation}
with
\mbox{$
{\cal B}_0=
({A_0}/{4R_0})
(h_0/R_0)^2
$}
setting the characteristic magnetic field scale. Using Eq.~\eqref{eq:intro_OAM_current}, the TPC density of an LG beam is purely azimuthal, $\mathbfcal{J}^{(\ell)}(\bm R)
    =\ell |u_\ell(\bm R)|^2 {\hat{\bm\phi}}/{R} $.
Substituting this into the OAM-projected GOT gives
\begin{equation}
    {\cal B}^{(\ell)}_{ab}
    =
    \ell
    \int_{\bm R}
    |u_\ell(\bm R)|^2
    \frac{{\cal A}_a(\bm R)\hat\phi_b}{R}.
    \label{eq:Lambda_LG_overlap}
\end{equation}
Thus \({\cal B}_{ab}^{(\ell)}\) is OAM-odd,
\({\cal B}^{(-\ell)}_{ab}=-{\cal B}^{(\ell)}_{ab}\), and has the units of an effective magnetic field scale. For weak displacement, we expand
$
{\cal A}^{G}_a(\bm R-\bm d)
\approx
{\cal A}^{G}_a(\bm R)
-d_c\partial_c{\cal A}^{G}_a(\bm R)+{\cal O}(d^2)
$.
Angular symmetry makes the zeroth-order contribution vanish, leaving the linear
term~\cite{sm}, 
\begin{equation}
    \mathbfcal{B}^{(\ell)}
    \approx
    \frac{\ell \, {\cal B}_0}{R_0} 
    \frac{2^{|\ell|}}{(2+w^2_0/R^2_0)^{|\ell|+1}}
    \begin{pmatrix}
        d_y & d_x \\
        d_x & -d_y
    \end{pmatrix}.
    \label{eq:B_ab}
\end{equation}
Fig.~\ref{fig2}(a) shows the OAM-projected GOT through
the diagonal anisotropy
\(\mbox{${\cal B}^{(\ell)}_D={\cal B}^{(\ell)}_{xx}
-{\cal B}^{(\ell)}_{yy}$}\). The main panel shows
\(\mbox{${\cal B}^{(1)}_D$}\) as a function of the common displacement
\(\mbox{$d_x=d_y=d$}\), for \(\mbox{$\ell=1$}\) and
\(\mbox{$w_0=R_0$}\). For this symmetric displacement direction, the
off-diagonal symmetric channel
\(\mbox{${\cal B}^{(\ell)}_S={\cal B}^{(\ell)}_{xy}
+{\cal B}^{(\ell)}_{yx}$}\) is numerically identical to
\(\mbox{${\cal B}^{(\ell)}_D$}\). The response changes sign under reversal of
the displacement and reaches its largest magnitude at intermediate
\(\mbox{$d/R_0$}\), where the transverse phase current has substantial overlap with the displaced pseudo-gauge texture. It is suppressed at large \(\mbox{$|d|/R_0$}\), where the transverse phase current and pseudo-gauge texture no longer overlap.

The upper inset shows the dependence on the beam waist \(\mbox{$w_0/R_0$}\),
evaluated at \(\mbox{$\ell=1$}\) and \(\mbox{$d_x=d_y=0.1R_0$}\). The overlap
decreases as the optical mode becomes wider than the pseudo-gauge texture, reflecting
the reduced spatial matching between the vortex TPC and the pseudo-gauge
texture. The lower inset shows the OAM-index dependence
\(\mbox{${\cal B}^{(\ell)}_D$}\) for \(\mbox{$w_0=R_0$}\) and
\(\mbox{$d_x=d_y=0.1R_0$}\). The nonmonotonic behavior follows from the
competition between the explicit OAM factor in the TPC and the
outward shift of the LG intensity profile with increasing \(\ell\), as evident in Eq. \eqref{eq:B_ab}, which eventually reduces its overlap with the localized pseudo-gauge texture.

\PRLsection{Nonlocal elasto-optic response}
We model the electronic structure near valley \(\tau=\pm\) by a tilted massive
Dirac Hamiltonian
\begin{equation}
    {\cal H}_\tau(\bm p,\bm r,t)
    =
    \tau\bm\alpha\cdot\bm p
    +
    v\left(\bm p+e\bm A+e\tau\mathbfcal A\right)\cdot\bm\sigma_\tau
    +
    m_\tau\sigma_z ,
    \label{eq:Dirac}
\end{equation}
where \(\bm\sigma_\tau=(\tau\sigma_x,\sigma_y)\),
\(\bm A(\bm r,t)\) is the optical vector potential of the vortex light, and
\(\tau\mathbfcal A(\bm r)\) is the strain-induced valley-odd pseudo-gauge
field. The mass \(m_\tau=m_S+\tau m_H\) includes Semenoff~\cite{Semenoff1984} and Haldane~\cite{Haldane1988}
components, while the tilt vector \(\bm\alpha=\alpha(\cos\vartheta,\sin\vartheta)\)
breaks the continuous rotational symmetry of the Dirac cone.
The leading strain-assisted nonlocal response follows 
\begin{equation}
    j_i(\bm q,\omega)
    =
    \int_{\bm q'}
    \zeta_{ijab}(\omega)
    Q_b E_j(\bm q',\omega)
    {\cal A}_a(\bm k).
    \label{eq:second_order_kernel}
\end{equation}
As shown explicitly in the Supplemental Material~\cite{sm}, an untilted massive
Dirac cone with broken valley symmetry gives only an isotropic response,
\(\Lambda^{(s)}_{ab}=\delta_{ab}\Lambda^{(s)}_0\). For a circular bubble this contribution is not OAM-odd; a tilt is required to break continuous rotational symmetry and produce an OAM-odd signal. The valley-odd pseudo-gauge coupling also cancels between equivalent valleys, so valley inequivalence is required, here from coexisting Semenoff and Haldane masses or a valley-polarized chemical potential. The full elasto-optic response reads~\cite{sm} 
\begin{align}\label{eq:Lambda_ab}
\begin{pmatrix}
    \Lambda^{(s)}_{xx}-\Lambda^{(s)}_{yy}
    \\
\Lambda^{(s)}_{xy}+\Lambda^{(s)}_{yx}
\end{pmatrix}
     =
     \alpha^2 Z_s
     \begin{pmatrix}
         \cos 2\vartheta
         \\
         \sin 2\vartheta
     \end{pmatrix} + {\cal O}(\alpha^4),
\end{align}
where \mbox{$Z_s =\sum_\tau (Z_\tau+sZ'_\tau)$}, with 
\begin{align}
   Z_\tau
   &\!=\!
   -g_s\frac{i e^3}{4\pi}
    \tau
   \left[
        \frac{1+3y_\tau^2}{\Omega^2}
        +
        \frac{x_\tau f(x_\tau,y_\tau)}
        {8\mu_\tau^2(x_\tau^2-4)^4}
   \right]
   T(y_\tau),
   \\
   Z'_\tau
   &\!=\!
   -g_s\frac{i e^3}{4\pi}
    \tau
   \frac{
        y_\tau g(x_\tau,y_\tau)
   }{
        \mu_\tau^2(x_\tau^2-4)^4
   }
   T(y_\tau).
\end{align}
Here $g_s=2$ is the electron spin-degeneracy factor, $\Omega=\hbar\omega+i\gamma$, $x_\tau=\Omega/\mu_\tau$, $y_\tau=m_\tau/\mu_\tau$, and $T(y)=\Theta(1-|y|)(1-y^2)$. The resonance occurs at
\(\hbar\omega=2\mu_\tau\). In \(Z\), the first term is intraband,
whereas the second is interband; the spin-dependent term \(Z'\)
is purely interband. The explicit forms of \(f(x,y)\) and \(g(x,y)\) are
given in the Supplemental Material~\cite{sm}. The valley sum is finite only when valley symmetry is broken, through either coexisting Semenoff and Haldane masses, $|m_+|\neq |m_-|$, or valley polarization, $\mu_+\neq\mu_-$.
Using Eqs.~\eqref{eq:intro_OAM_overlap}, \eqref{eq:B_ab}, and
\eqref{eq:Lambda_ab}, the OAMD contrast for a displaced Gaussian bubble in
tilted massive Dirac electrons is
\begin{align}
\Delta \sigma^{\rm OAM}_{\ell,s}
\approx \frac{2^{|\ell|} \alpha^2 {\cal B}_0  {\rm Re}\!\left[\ell Z_s\right]}{(2+w^2_0/R^2_0)^{|\ell|+1}}
    \frac{d_x\sin 2\vartheta+d_y\cos 2\vartheta}{R_0}.
\end{align}
The signal is therefore controlled by the tilt magnitude and direction, the
bubble displacement, and the frequency dependence of \(Z_s\). Fig.~\ref{fig2}(b) shows the material coefficients entering the OAM-odd
response. The main panel plots the
interband response \({\rm Re}\,Z^{\rm inter}\) and the
SAM-resolved coefficient \({\rm Re}\,Z'\) near the interband resonance
\(\hbar\omega\simeq2\mu\), for \(\mu_+=\mu_-=\mu\) and
\(m_\tau=m_S+\tau m_H\). The strong resonant enhancement reflects the
fourth-order interband pole, broadened by finite \(\gamma\). The upper-left inset shows the corresponding single-valley \(\tau=+\)
interband response. Its larger amplitude reveals the partial cancellation
between time-reversed valleys in the valley-summed curves. The lower-right
inset compares the low-frequency intraband response with its \(\tau=+\)
component, showing the Drude-like contribution present in \(Z\) but absent
from the purely interband coefficient \(Z'\).

\PRLsection{Conclusion and outlook}
We show that OAM-odd absorption requires both spatial dispersion and inhomogeneity. Together, they enable transverse momentum transfer to the electronic system when continuous translational symmetry is broken. This no-go theorem is specific to spatially integrated direct optical absorption. Related momentum-conservation constraints may also apply to photoemission, but need not extend to photoluminescence, which depends on post-excitation dynamics, or to nonlinear photocurrents, whose terminal readout samples selected positions and directions.

In the strained Dirac model considered here, the valley-odd pseudo-gauge coupling imposes an
additional constraint on valley-summed detection: inequivalent valleys are
required to avoid cancellation. This constraint is not generic. Even in a
time-reversal-symmetric system, valley-resolved OAM dichroism is allowed
through \(\ell Z_\tau\) and \(\ell s Z'_\tau\). These couplings imply that opposite vortex helicities address opposite
valleys, closely analogous to SAM-based valley dichroism in inversion-broken
massive Dirac systems~\cite{Yao2008,Cao2012,Mak2012}.
Promising platforms include graphene nanobubbles with large pseudo-magnetic fields~\cite{Levy2010PseudoLandau,Jiang2017VisualizingPMF,Nigge2019RoomTemperaturePLL,Khestanova2016UniversalBubbles}, strained transition-metal dichalcogenides with valley-selective optics~\cite{CastellanosGomez2013LocalStrain,Blundo2021StrainTuning,Rostami2015TheoryStrainTMD,Rostami2018PiezoelectricityValleyChern,Zhang2023OAMEmitter}, twisted moir\'e bilayers with reconstruction-induced textures~\cite{Naik2018MoireTMD,Wu2019TopologicalTMDMoire,Regan2020MottWSe2Moire,Tang2020SimulationHubbardMoire}, and anisotropic Dirac or inversion-broken black-phosphorus~\cite{Rostami2020,Low2015BlackPhosphorus}.

OAM-carrying beams have been realized across the electromagnetic spectrum,
from RF/microwave and THz frequencies to the visible, EUV, and soft-X-ray
regimes
~\cite{Ji2020,Fanciulli2022,Sirenko2021TerahertzVortex,
Petrov2022THzVortexReview,Gauthier2017TunableOrbitalAngularMomentum}.
For an order-of-magnitude estimate, we take a bubble with
\(\mbox{$h_0/R_0\sim0.1$}\), as observed in 2D material
bubbles~\cite{Khestanova2016UniversalBubbles}, displaced by
\(\mbox{$d_x=d_y\sim0.1R_0$}\), with
\(\mbox{$R_0\sim1\,\mu{\rm m}$}\). For \(\mbox{$\ell=1$}\) and
\(\mbox{$w_0=R_0$}\), Eq.~\eqref{eq:B_ab} gives an OAM-projected geometric
overlap scale \(\mbox{${\cal B}\sim{\rm mT}$}\). Combining this with the
resonant material scale \(\mbox{$Z_s\sim10^5 e^3/(2\pi\mu^2)$}\) at
\(\mbox{$\mu\sim0.1\,{\rm eV}$}\) and a tilt velocity
\(\mbox{$\alpha\sim10^5\,{\rm m/s}$}\), we obtain
\(\mbox{$\Delta\sigma^{\rm OAM}_{\ell,s}\sim10^{-3}e^2/\hbar$}\).
The single-valley dichroism can exceed the valley-summed response because
time-reversed valleys partially cancel, as shown in Fig.~\ref{fig2}(b).

\begin{acknowledgments}
The author thanks Philip Hofmann, Mauro Fanciulli, Søren Ulstrup, Emmanuele Cappelluti, and Weibo Gao for helpful discussions. This work was supported by EPSRC Grant No.~UKRI122 and Royal Society Grant Nos.~IES\textbackslash R2\textbackslash 242309 and RGS\textbackslash R2\textbackslash 252194.
\end{acknowledgments}

\bibliography{ref.bib}

\begin{thebibliography}{65}
\bibitem [{LGG()}]{LGGlobalPhase}%
  \BibitemOpen
  \href@noop {} {}\bibinfo {note} {The Fourier transform of the
  Laguerre--Gaussian envelope contains an \(\ell\)-dependent global phase,
  whose precise form depends on the Fourier-transform and angular-phase
  convention. This global phase has no effect on the dichroic signal, which is
  quadratic in the optical-mode amplitude, and is omitted.}\BibitemShut {Stop}%
\end{thebibliography}

\end{document}


\title{Supplemental Material for ``Nonlocal Orbital-Angular-Momentum Dichroism of Vortex Light in Strained Crystals''}

\author{Habib Rostami}
\email{hr745@bath.ac.uk}
\affiliation{Department of Physics, University of Bath, Claverton Down, Bath BA2 7AY, United Kingdom}

\date{\today}

\maketitle
\tableofcontents 
\section{Vortex-light phase current and strain-field overlap: Eq.~(18) of main text}

Substituting Eq.~(17) into Eq.~(16) of the main text yields 
\begin{align}
    \delta\Sigma_{\ell, s}(\omega)
    =
    \Lambda^{(s)}_{ab}(\omega)
    \int_{\bm q}\int_{\bm q'}
    \tilde u^\ast_\ell(\bm q)\tilde u_\ell(\bm q')
    {\cal A}_a(\bm k) Q_b ,
    \quad \text{with} \quad
    \bm Q=\frac{\bm q+\bm q'}{2},
    \quad 
    \bm k=\bm q-\bm q',
    \label{eq:S1_deltaSigma_nonlocal}
\end{align}
and
\begin{align}\label{sm_eq:Lambda}
    \Lambda^{(s)}_{ab}(\omega)
 =e^\ast_{s,i} \zeta_{ijab}(\omega) e_{s,j}.  
 \end{align}
Here and below, repeated Cartesian indices are summed. In this section we show that the momentum-space overlap appearing in Eq.~\eqref{eq:S1_deltaSigma_nonlocal} can be written as a real-space overlap between the strain-induced pseudo-gauge field and the phase current of the optical vortex.
We use the Fourier convention
\begin{equation}
    {\cal A}_a(\bm k)
    =
    \int d^2R\,
    {\cal A}_a(\bm R)e^{-i\bm k\cdot\bm R},
    \qquad
    \int_{\bm q}\equiv \int\frac{d^2q}{(2\pi)^2}.
    \label{eq:S1_fourier_convention}
\end{equation}
For a scalar optical mode,
\begin{equation}
    u_\ell(\bm R)
    =
    \int_{\bm q}
    \tilde u_\ell(\bm q)e^{i\bm q\cdot\bm R},
    \qquad
    u_\ell^*(\bm R)
    =
    \int_{\bm q}
    \tilde u_\ell^*(\bm q)e^{-i\bm q\cdot\bm R}.
    \label{eq:S1_mode_fourier}
\end{equation}
The leading OAM-sensitive tensor is therefore
\begin{equation}
    {\cal B}^{(\ell)}_{ab}
    =
    \int_{\bm q}\int_{\bm q'}
    \tilde u_\ell^*(\bm q)\tilde u_\ell(\bm q')
    {\cal A}_a(\bm q-\bm q')
    \frac{q_b+q'_b}{2}.
    \label{eq:S1_Lambda_momentum}
\end{equation}
The factor \({\cal A}_a(\bm q-\bm q')\) carries the transverse momentum transferred by the strain texture. Substituting Eq.~\eqref{eq:S1_fourier_convention} into Eq.~\eqref{eq:S1_Lambda_momentum} gives
\begin{align}
    {\cal B}^{(\ell)}_{ab}
    =
    \int d^2R\,{\cal A}_a(\bm R)
    \int_{\bm q}\int_{\bm q'}
    \tilde u_\ell^*(\bm q)\tilde u_\ell(\bm q')
    e^{-i(\bm q-\bm q')\cdot\bm R}
    \frac{q_b+q'_b}{2}.
    \label{eq:S1_Lambda_intermediate}
\end{align}
We now show that the second factor in Eq.~\eqref{eq:S1_Lambda_intermediate} is precisely the phase-current density of the optical mode.
From Eq.~\eqref{eq:S1_mode_fourier}, the derivatives of the mode and its complex conjugate are
\begin{align}
    \partial_b u_\ell(\bm R)
    &=
    \int_{\bm q'}
    i q'_b
    \tilde u_\ell(\bm q')
    e^{i\bm q'\cdot\bm R},
    \nonumber\\
    \partial_b u_\ell^*(\bm R)
    &=
    \int_{\bm q}
    (-i q_b)
    \tilde u_\ell^*(\bm q)
    e^{-i\bm q\cdot\bm R}.
    \label{eq:S1_mode_derivatives}
\end{align}
It follows that
\begin{align}
    u_\ell^*(\bm R)\partial_b u_\ell(\bm R)
    &=
    \int_{\bm q}\int_{\bm q'}
    \tilde u_\ell^*(\bm q)\tilde u_\ell(\bm q')
    e^{-i(\bm q-\bm q')\cdot\bm R}
    i q'_b ,
    \label{eq:S1_u_star_grad_u}
    \\
    u_\ell(\bm R)\partial_b u_\ell^*(\bm R)
    &=
    \int_{\bm q}\int_{\bm q'}
    \tilde u_\ell^*(\bm q)\tilde u_\ell(\bm q')
    e^{-i(\bm q-\bm q')\cdot\bm R}
    (-i q_b).
    \label{eq:S1_u_grad_u_star}
\end{align}
Subtracting Eq.~\eqref{eq:S1_u_grad_u_star} from Eq.~\eqref{eq:S1_u_star_grad_u} gives
\begin{align}
    \frac{1}{2i}
    \left[
        u_\ell^*(\bm R)\partial_b u_\ell(\bm R)
        -
        u_\ell(\bm R)\partial_b u_\ell^*(\bm R)
    \right]
    =
    \int_{\bm q}\int_{\bm q'}
    \tilde u_\ell^*(\bm q)\tilde u_\ell(\bm q')
    e^{-i(\bm q-\bm q')\cdot\bm R}
    \frac{q_b+q'_b}{2}.
    \label{eq:S1_phase_current_identity}
\end{align}
This proves the desired identity. We therefore define the transverse phase-current density
\begin{equation}
    {\cal J}^{(\ell)}_{b}(\bm R)
    =
    \frac{1}{2i}
    \left[
        u_\ell^*(\bm R)\partial_b u_\ell(\bm R)
        -
        u_\ell(\bm R)\partial_b u_\ell^*(\bm R)
    \right]
    =
    {\rm Im}\left[
        u_\ell^*(\bm R)\partial_b u_\ell(\bm R)
    \right].
    \label{eq:S1_phase_current_def}
\end{equation}
Combining Eqs.~\eqref{eq:S1_Lambda_intermediate} and \eqref{eq:S1_phase_current_identity}, the momentum-space tensor in Eq.~\eqref{eq:S1_Lambda_momentum} becomes the real-space geometric overlap
\begin{equation}
    {\cal B}^{(\ell)}_{ab}
    =
    \int d^2R\,
    {\cal A}_a(\bm R)
    {\cal J}^{(\ell)}_{b}(\bm R).
    \label{eq:S1_Lambda_real_space}
\end{equation}
Equation~\eqref{eq:S1_Lambda_real_space} shows that \({\cal B}^{(\ell)}_{ab}\) measures the overlap between the strain-induced pseudo-gauge field component \({\cal A}_a\) and the local phase flow of the vortex beam along direction \(b\).
For a scalar LG beam with the same radial envelope for opposite OAM, \(u_{-\ell}=u_\ell^*\). Equation~\eqref{eq:S1_phase_current_def} then gives
\begin{equation}
    {\cal J}^{(-\ell)}_{b}(\bm R)
    =
    -
    {\cal J}^{(\ell)}_{b}(\bm R).
    \label{eq:S1_current_OAM_odd}
\end{equation}
Using Eq.~\eqref{eq:S1_Lambda_real_space}, this immediately implies
\begin{equation}
    {\cal B}^{(-\ell)}_{ab}
    =
    -
    {\cal B}^{(\ell)}_{ab}.
    \label{eq:S1_Lambda_OAM_odd}
\end{equation}
The leading OAM-odd correction to the mode-projected response can therefore be written as
\begin{equation}
    \delta\Sigma_{\ell, s}(\omega)
    =
    {\cal B}^{(\ell)}_{ab}\Lambda^{(s)}_{ab}(\omega).
    \label{eq:S1_deltaSigma_Lambda}
\end{equation}

\section{Displaced Gaussian bubble contribution: Eq.~(21) of main text}

We evaluate the geometric overlap tensor
\begin{align}
    {\cal B}^{(\ell)}_{ab}
    =
    \int d^2R\,
    {\cal A}^{G}_a(\bm R-\bm d)
    {\cal J}^{(\ell)}_b(\bm R),
    \label{eq:B_def_displaced_SM}
\end{align}
where \(\bm d=(d_x,d_y)\) is the displacement of the Gaussian bubble relative to the optical-vortex core. For \(|\bm d|\) small compared with the beam waist and bubble radius,
\begin{align}
    {\cal A}^{G}_a(\bm R-\bm d)
    =
    {\cal A}^{G}_a(\bm R)
    -
    d_c\partial_c{\cal A}^{G}_a(\bm R)
    +
    {\cal O}(d^2),
\end{align}
so that
\begin{align}
    {\cal B}^{(\ell)}_{ab}
    =
    \int d^2R\,
    {\cal A}^{G}_a(\bm R)
    {\cal J}^{(\ell)}_b(\bm R)
    -
    d_c
    \int d^2R\,
    \partial_c{\cal A}^{G}_a(\bm R)
    {\cal J}^{(\ell)}_b(\bm R)
    +
    {\cal O}(d^2).
    \label{eq:B_expansion_displacement_SM}
\end{align}

For the centered circular Gaussian bubble we use the convention
\begin{align}
    {\cal A}^G_x(\bm R)
    =
    {\cal A}^G(R)\cos2\phi,
    \qquad
    {\cal A}^G_y(\bm R)
    =
    -{\cal A}^G(R)\sin2\phi,
    \label{eq:AG_components_SM}
\end{align}
with radial profile
\begin{align}
    {\cal A}^G(R)
    =
    R_0{\cal B}_0
    \frac{
    [1+(R/R_0)^2]e^{-R^2/R_0^2}-1
    }{
    (R/R_0)^2
    }.
    \label{eq:gaussian_A_profile_SM}
\end{align}
The transverse phase-current density of the lowest-radial-index LG mode is
\begin{align}
    {\cal J}^{(\ell)}_b(\bm R)
    =
    {\rm Im}\,[u_\ell^*(\bm R)\partial_b u_\ell(\bm R)],
\end{align}
or explicitly
\begin{align}
    {\cal J}^{(\ell)}_x(\bm R)
    =
    -\ell |u_\ell(R)|^2\frac{\sin\phi}{R},
    \qquad
    {\cal J}^{(\ell)}_y(\bm R)
    =
    \ell |u_\ell(R)|^2\frac{\cos\phi}{R}.
    \label{eq:LG_current_SM}
\end{align}
Here
\begin{align}
    |u_\ell(R)|^2
    =
    \frac{2^{|\ell|+1}}{\pi |\ell|! w_0^2}
    \left(\frac{R}{w_0}\right)^{2|\ell|}
    e^{-2R^2/w_0^2}.
    \label{eq:vortex_density_SM}
\end{align}

The zeroth-order contribution in Eq.~\eqref{eq:B_expansion_displacement_SM} vanishes by angular symmetry. Indeed,
\begin{align}
\int d^2R\,
{\cal A}^{G}_a(\bm R)
{\cal J}^{(\ell)}_b(\bm R)
&=
\ell
\int_0^\infty dR\,
{\cal A}^G(R)|u_\ell(R)|^2
\nonumber\\
&\quad\times
\int_0^{2\pi}d\phi
\begin{pmatrix}
-\cos2\phi\sin\phi
&
\cos2\phi\cos\phi
\\
\sin2\phi\sin\phi
&
-\sin2\phi\cos\phi
\end{pmatrix}_{ab}
=0 ,
\label{eq:zeroth_order_vanishes_SM}
\end{align}
where the Jacobian \(R\) in \(d^2R=R\,dR\,d\phi\) cancels the \(1/R\) factor in Eq.~\eqref{eq:LG_current_SM}.
We therefore keep only the linear term in \(\bm d\). The Cartesian derivatives of Eq.~\eqref{eq:AG_components_SM} are
\begin{align}
\partial_c{\cal A}^G_a
=
\begin{pmatrix}
\partial_x{\cal A}^G_x & \partial_x{\cal A}^G_y \\[4pt]
\partial_y{\cal A}^G_x & \partial_y{\cal A}^G_y
\end{pmatrix}_{ca}
=
\begin{pmatrix}
{\cal A}^{G\prime}\cos\phi\cos2\phi
+\dfrac{2{\cal A}^G}{R}\sin\phi\sin2\phi
&
-{\cal A}^{G\prime}\cos\phi\sin2\phi
+\dfrac{2{\cal A}^G}{R}\sin\phi\cos2\phi
\\[7pt]
{\cal A}^{G\prime}\sin\phi\cos2\phi
-\dfrac{2{\cal A}^G}{R}\cos\phi\sin2\phi
&
-{\cal A}^{G\prime}\sin\phi\sin2\phi
-\dfrac{2{\cal A}^G}{R}\cos\phi\cos2\phi
\end{pmatrix}_{ca},
\label{eq:cartesian_grad_gaussian_gauge_SM}
\end{align}
where \({\cal A}^{G\prime}=d{\cal A}^G/dR\). The angular averages needed below are
\begin{align}
    \int_0^{2\pi}d\phi\,
    \cos\phi\,
    \partial_c{\cal A}^G_a
    =
    \frac{\pi}{2}
    \left(
    {\cal A}^{G\prime}
    +
    2\frac{{\cal A}^G}{R}
    \right)
    \begin{pmatrix}
    1&0\\
    0&-1
    \end{pmatrix}_{ca},
    \label{eq:angular_cos_SM}
\end{align}
and
\begin{align}
    \int_0^{2\pi}d\phi\,
    \sin\phi\,
    \partial_c{\cal A}^G_a
    =
    -\frac{\pi}{2}
    \left(
    {\cal A}^{G\prime}
    +
    2\frac{{\cal A}^G}{R}
    \right)
    \begin{pmatrix}
    0&1\\
    1&0
    \end{pmatrix}_{ca}.
    \label{eq:angular_sin_SM}
\end{align}

It is useful to define the radial scale
\begin{align}
    {\cal I}_\ell
    =
    -\frac{\pi\ell}{2}
    \int_0^\infty dR\,
    |u_\ell(R)|^2
    \left[
    {\cal A}^{G\prime}(R)
    +
    2\frac{{\cal A}^G(R)}{R}
    \right].
    \label{eq:Iell_definition_SM}
\end{align}
With this convention, Eqs.~\eqref{eq:angular_cos_SM} and \eqref{eq:angular_sin_SM} give
\begin{align}
    \int d^2R\,
    \partial_c{\cal A}^G_a(\bm R)
    {\cal J}^{(\ell)}_x(\bm R)
    =
    -{\cal I}_\ell
    \begin{pmatrix}
    0&1\\
    1&0
    \end{pmatrix}_{ca},
    \label{eq:grad_overlap_Jx_SM}
\end{align}
and
\begin{align}
    \int d^2R\,
    \partial_c{\cal A}^G_a(\bm R)
    {\cal J}^{(\ell)}_y(\bm R)
    =
    -{\cal I}_\ell
    \begin{pmatrix}
    1&0\\
    0&-1
    \end{pmatrix}_{ca}.
    \label{eq:grad_overlap_Jy_SM}
\end{align}
Contracting Eqs.~\eqref{eq:grad_overlap_Jx_SM} and \eqref{eq:grad_overlap_Jy_SM} with \(d_c\), and keeping the overall minus sign in Eq.~\eqref{eq:B_expansion_displacement_SM}, yields
\begin{align}
    {\cal B}^{(\ell)}
    =
    {\cal I}_\ell
    \begin{pmatrix}
    d_y & d_x\\
    d_x & -d_y
    \end{pmatrix}
    +
    {\cal O}(d^2).
    \label{eq:B_final_displaced_bubble_SM}
\end{align}

For the Gaussian profile in Eq.~\eqref{eq:gaussian_A_profile_SM}, the radial kernel simplifies to
\begin{align}
    {\cal A}^{G\prime}(R)
    +
    2\frac{{\cal A}^G(R)}{R}
    =
    -2{\cal B}_0
    \frac{R}{R_0}
    e^{-R^2/R_0^2}.
    \label{eq:radial_kernel_simplified_SM}
\end{align}
Thus
\begin{align}
    {\cal I}_\ell
    =
    \pi\ell
    \frac{{\cal B}_0}{R_0}
    \int_0^\infty R\,dR\,
    |u_\ell(R)|^2
    e^{-R^2/R_0^2}.
    \label{eq:Iell_simplified_SM}
\end{align}
Substituting Eq.~\eqref{eq:vortex_density_SM}, one obtains
\begin{align}
    {\cal I}_\ell
    =
    \ell\,
    {\cal B}_0
    \frac{ 2^{|\ell|}
    R_0^{2|\ell|+1}
    }{
    \left(w_0^2+2R_0^2\right)^{|\ell|+1}
    }.
    \label{eq:Iell_explicit_SM}
\end{align}
The overlap is therefore odd in both \(\ell\) and \(\bm d\), and vanishes for a coaxial circular bubble.
 
\section{Nonlocal elasto-optic coefficients: Eqs. (24)–(26) of the main text}

We first evaluate the interband response using a diagrammatic Kubo approach, showing that the filled-valence-band contribution vanishes while tilt-dependent Pauli blocking generates a finite Fermi-surface contribution. We then derive the intraband contribution using Boltzmann kinetic theory.

\subsection{Kubo analysis of the interband contribution}

\begin{figure}[b]
\centering
\begin{tikzpicture}[
    >=Latex,
    lab/.style={font=\scriptsize},
    fermion/.style={
        line width=0.72pt,
        postaction={
            decorate,
            decoration={
                markings,
                mark=at position 0.55 with {
                    \arrow{Latex[length=1.75mm,width=1.15mm]}
                }
            }
        }
    },
    openvtx/.style={
        circle,
        draw=black,
        fill=white,
        line width=0.65pt,
        inner sep=1.45pt
    },
    solidvtx/.style={
        circle,
        fill=black,
        inner sep=1.55pt
    },
    squarevtx/.style={
        rectangle,
        fill=black,
        minimum size=4.1pt,
        inner sep=0pt
    }
]


\coordinate (T) at (0.00,1.35);      
\coordinate (R) at (1.55,-0.85);     
\coordinate (L) at (-1.55,-0.85);    


\draw[fermion]
    (T) --
    node[pos=0.35, above right=2pt, rotate=-55, lab]
    {$(n,\bm p)$}
    (R);

\draw[fermion]
    (R) --
    node[pos=0.50, below=3pt, lab]
    {$(n+m,\bm p+\hbar\bm q')$}
    (L);

\draw[fermion]
    (L) --
    node[pos=0.85, above left=2pt, rotate=55, lab]
    {$(n+m,\bm p+\hbar\bm q)$}
    (T);


\node[openvtx] at (T) {};
\node[solidvtx] at (R) {};
\node[squarevtx] at (L) {};

\node[lab, above=3pt] at (T) {$j_i$};
\node[lab, below right=2pt] at (R) {$j_j$};
\node[lab, below left=2pt] at (L) {$\lambda_a$};

\end{tikzpicture}
\caption{
Triangle diagram for the nonlocal elasto-optic response function.
The open circle denotes the output-current vertex $j_i$, the solid circle
denotes the optical-current vertex $j_j$, and the solid square denotes the
pseudo-gauge strain vertex $\lambda_a$. The indices $n$ and $m$ denote,
respectively, the internal fermionic Matsubara frequency and the external
bosonic Matsubara frequency. The wave vector $\bm q'$ is carried by the
incoming photon, while $\bm k$ is supplied by the static strain field; hence
the generated current carries wave vector $\bm q=\bm q'+\bm k$.
}
\label{fig:triangle-kubo}
\end{figure}
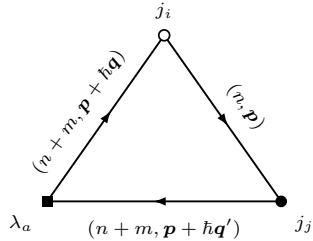

To fix conventions, consider the generic second-order response
\begin{align}\label{eq:second_order_kernel}
  J_i(\bm q_\Sigma,\omega_\Sigma)
  &=
  \sum_{\omega_1,\omega_2}
  \sum_{\bm q_1,\bm q_2}
  \Gamma_{ija}(\bm q_1,\omega_1;\bm q_2,\omega_2)\,
  A_j(\bm q_1,\omega_1)
  {\cal A}_a(\bm q_2,\omega_2)
  \delta_{\bm q_\Sigma,\bm q_1+\bm q_2}
\delta_{\omega_\Sigma,\omega_1+\omega_2}.
\end{align}
The triangle diagram shown in Fig.~\ref{fig:triangle-kubo} gives, in Matsubara notation,
\begin{align}\label{eq:triangle_kernel} 
\Gamma_{ija}(\bm q_1,m_1;\bm q_2,m_2)
=
-\frac{1}{\beta}
\sum_n
\int_{\bm p}
\mathrm{tr}\!
\Big[
j_i G(n,\bm p)\,
j_j 
G(n+m_1,\bm p+\hbar\bm q_1)\,
\lambda_a G(n+m_1+m_2,\bm p+\hbar\bm q_1+\hbar\bm q_2)
\Big]. 
\end{align}
with $\beta=1/k_BT$ and the Green's function
\begin{equation}
    G(n,\bm p)=\left [n+\mu_\tau-{\cal H}_\tau(\bm p)\right ]^{-1}.
\end{equation}
Here $\mu_\tau$ is the valley-resolved chemical potential, \(n\equiv i\omega_n\) denotes a fermionic Matsubara frequency, while
\(m_{1,2}\equiv i\Omega_{m_{1,2}}\) denote bosonic Matsubara frequencies.
The internal variable \(\bm p\) is the electronic momentum, whereas
\(\bm q_1\) and \(\bm q_2\) are external wave vectors. Accordingly, external
momentum transfer enters the electronic Green's functions as
\(\hbar\bm q_1\) and \(\hbar\bm q_2\).
The optical vertices are
$(j_x,j_y)=-e\left(u_x+\tau v\sigma_x,\;u_y+v\sigma_y\right)$,
whereas the pseudo-gauge-field vertices are $(\lambda_x,\lambda_y)=
-ev\left(\sigma_x,\;\tau\sigma_y\right)$. The factor $\tau$ encodes the valley-odd coupling of strain. The trace is over the pseudospin degree of freedom.  Comparison with Eq.~\eqref{eq:second_order_kernel} gives
$\bm q_\Sigma=\bm q$, $\omega_\Sigma=\omega$, $\bm q_1=\bm q'$, and $\omega_1=\omega$, while the static strain field carries $\bm q_2=\bm k=\bm q-\bm q'$ and $\omega_2=0$. Thus the strain texture supplies the missing transverse momentum that mixes
incoming optical momentum $\bm q'$ into outgoing momentum $\bm q$. Therefore, after considering permutation symmetry we have  
\begin{align}
 K_{ija} \left(\bm Q,\bm k;\omega\right)
 &=\frac{1}{2}\Big\{\Gamma_{ija} \Big(\bm Q-{\bm k}/{2},\omega; \bm k,0) +
 \Gamma_{iaj} (\bm k,0;\bm Q-{\bm k}/{2},\omega)\Big\}.
\end{align}
We now expand the symmetrized kernel to first order in the average external
wave vector \(\bm Q\). Since the external wave vector enters the Green's
function as \(\hbar\bm Q\), the expansion produces the factor
\(\hbar Q_b\partial_{p_b}G\). Using
\begin{equation}
\partial_{p_b}G
=
-G\partial_{p_b}G^{-1}G,
\qquad
\partial_{p_b}G^{-1}
=
-\partial_{p_b}{\cal H}_\tau
=
\frac{j_b}{e},
\end{equation}
one obtains
\begin{equation}
\partial_{p_b}G
=
-\frac{1}{e}G j_b G .
\end{equation}
The minus sign from this derivative cancels the overall fermion-loop sign in
\(\Gamma\). Thus, after setting \(\bm k=0\), we have three possible contributions at each valley point. Expanding the symmetrized second-order kernel to linear order in the average
optical wave vector $\bm Q$, and then setting $\bm k=0$, gives \begin{align}
K^{(1)}_{ijab}(m)
=
\frac{\hbar}{2\beta e}
\sum_n
\int_{\bm p}
\chi_{ijab}(n,m),
\end{align} 
with 
\begin{align}\label{eq:chi_ijab}
\chi_{ijab}& =
{\rm Tr} [j_i G(n) j_j G(n+m) j_b G(n+m) \lambda_a G(n+m )] 
\nonumber \\&+ 
{\rm Tr} [j_i G(n) j_j G(n+m) \lambda_a G(n+m ) j_b G(n+m)]
\nonumber\\&+ 
{\rm Tr} [j_i G(n) \lambda_a G(n) j_j G(n+m)j_b G(n+m)].
\end{align}
After performing the momentum integral and Matsubara summation, we obtain
\(K_{ijab}(\omega)\equiv K_{ijab}(m\to \Omega=\hbar\omega+i\gamma)\), where
\(\gamma\to0^+\) in the clean limit. A finite \(\gamma\) may be retained
phenomenologically to model relaxation. The response tensor is then
\begin{equation}
    \zeta_{ijab}(\omega)
    =
    \frac{K_{ijab}(\omega)-K_{ijab}(0)}{i\omega}.
\end{equation}
The division by \(i\omega\) converts the response to the vector potential into
the response to the electric field.
The static subtraction enforces gauge invariance: a uniform static vector
potential cannot generate a physical current. Using Eq.~\eqref{sm_eq:Lambda}, the spin-projected elasto-optic tensor is
\begin{align}
    \Lambda^{(s)}_{xx}
    &=
    \frac{\zeta_{xxxx}+\zeta_{yyxx}}{2}
    +is\,\frac{\zeta_{xyxx}-\zeta_{yxxx}}{2},
    \\
    \Lambda^{(s)}_{yy}
    &=
    \frac{\zeta_{xxyy}+\zeta_{yyyy}}{2}
    +is\,\frac{\zeta_{xyyy}-\zeta_{yxyy}}{2},
    \\
    \Lambda^{(s)}_{xy}
    &=
    \frac{\zeta_{xxxy}+\zeta_{yyxy}}{2}
    +is\,\frac{\zeta_{xyxy}-\zeta_{yxxy}}{2},
    \\
    \Lambda^{(s)}_{yx}
    &=
    \frac{\zeta_{xxyx}+\zeta_{yyyx}}{2}
    +is\,\frac{\zeta_{xyyx}-\zeta_{yxyx}}{2}.
\end{align}

We first apply this theory to a single-valley massive Dirac model without a
tilt term. After evaluating the trace and performing the Matsubara summation,
the response separates into occupation-proportional terms and Fermi-surface
terms proportional to derivatives of the occupation function. The
occupation-proportional part of the angular-averaged kernel is generally
nonzero locally in momentum space. However, after summing the three routed
permutations in Eq.~\eqref{eq:chi_ijab}, the radial integrand becomes a total
derivative in \(p\). For a fully occupied valence band, the integral therefore
reduces to a boundary term, which vanishes in the insulating regime. In the
metallic regime, however, a partially occupied conduction band produces a
finite boundary contribution at the Fermi surface, where the integrand is
generally nonzero. For the isotropic massive Dirac model without tilt, the explicit
calculation gives
\begin{align}
    \Lambda^{(s)}_{ab}
    &=\delta_{ab} \Lambda_{0}=\delta_{ab}\sum_\tau
    \tau \frac{e^3 v^2}{4\pi }
    \frac{
    i(\mu_\tau^2-m_\tau^2)
    \left[
        m_\tau^2(3\Omega^3-20\mu_\tau^2\Omega)
        -12\mu_\tau^4\Omega
        +\mu_\tau^2\Omega^3
        +s m_\tau(48\mu_\tau^4-4\mu_\tau^2\Omega^2)
    \right]
    }{
    8\mu_\tau^5(\Omega^2-4\mu_\tau^2)^2
    } ,
\end{align}
where \(\Omega=\hbar\omega+i\gamma\). Thus, valley-symmetry breaking in this
isotropic massive Dirac model, either through a Haldane mass
\(m_{+}-m_{-}=2m_H\) or through a valley chemical-potential imbalance
\(\mu_{+}\neq\mu_{-}\), makes the diagonal components of
\(\bm\Lambda\) finite. However, rotational symmetry still enforces $\Lambda^{(s)}_{xx}=\Lambda^{(s)}_{yy}$ and $\Lambda^{(s)}_{xy}=\Lambda^{(s)}_{yx}=0$.
Consequently, these contributions cannot generate OAM-odd absorption for a
circularly symmetric Gaussian bubble, for which the signal is controlled by the
anisotropic combinations
\(\Lambda^{(s)}_{xx}-\Lambda^{(s)}_{yy}\) and
\(\Lambda^{(s)}_{xy}+\Lambda^{(s)}_{yx}\).

We now include the tilt term, which breaks rotational symmetry and can therefore
generate a nonzero OAM-odd response. The tilted massive Dirac Hamiltonian is
\begin{equation}
    H_\tau(\bm p)
    =
    \tau \bm\alpha\cdot\bm p 
    +
    \tau v p_x\sigma_x
    +
    v p_y\sigma_y
    +
    m_\tau\sigma_z .
\end{equation}
Since the tilt term is proportional to \(\sigma_0\), it shifts both bands by
the same amount but does not modify the band eigenstates. The interband
transition energy is therefore unchanged,
\begin{equation}
    \varepsilon_{c,\tau}(\bm p)-\varepsilon_{v,\tau}(\bm p)
    =
    2E_{p\tau},
    \qquad
    E_{p\tau}=\sqrt{(vp)^2+m_\tau^2}.
\end{equation}
Moreover, the tilt contribution to the current vertex is also proportional to
\(\sigma_0\), and hence has vanishing interband matrix elements, $\langle c,\bm p|\sigma_0|v,\bm p\rangle=0$.
Thus, within the interband sector, the tilt changes neither the transition
energy nor the interband velocity matrix elements. Its only effect is through
the Pauli-blocking factor, $ f_{v,\tau}(\bm p)-f_{c,\tau}(\bm p)$,
which is modified by the tilt-induced deformation of the Fermi surface.

For electron doping, the valence band is fully occupied,
\(f_{v,\tau}=1\). The valence contribution is then identical to that of the
rotationally invariant untilted model and it vanishes. The remaining tilt-dependent contribution comes from
the partially occupied conduction band and can be written as
\begin{equation}
    \Lambda^{(s)}_{ab}(\omega)
    =
    \int_{\bm p}
    S^{(s)}_{ab}(\bm p,\omega)\,
    \Theta\!\left(
        \mu_\tau-\tau\bm\alpha\cdot\bm p-E_{p\tau}
    \right).
\end{equation}
where $S_{ab}(\bm p,\omega)$ is the untilted interband kernel. For weak tilt, the Pauli-blocking
factor may be expanded as
\begin{align}
    \Lambda^{(s)}_{ab}(\omega)
    &=
    \int_{\bm p}
    S^{(s)}_{ab}(\bm p,\omega)
    \bigg[
    \Theta\!\left(\mu_\tau-E_{p\tau}\right)
    -
    (\tau\bm\alpha\cdot\bm p)
    \frac{\partial}{\partial\mu_\tau}
    \Theta\!\left(\mu_\tau-E_{p\tau}\right)
    +
    \frac{1}{2}
    (\tau\bm\alpha\cdot\bm p)^2
    \frac{\partial^2}{\partial\mu_\tau^2}
    \Theta\!\left(\mu_\tau-E_{p\tau}\right)
    +\cdots
    \bigg].
\end{align}
The first term reproduces the result of the isotropic massive Dirac model and
therefore gives no OAM-odd signal for a circularly symmetric Gaussian bubble.
The term linear in the tilt vanishes after angular integration. Hence the
leading nonvanishing OAM-odd interband contribution is second order in
\(\alpha\):
\begin{align}
    \Lambda^{(s)}_{ab}(\omega)
    &=\delta_{ab}\Lambda^{(s)}_{0}(\omega) + 
    \frac{1}{2}
    \int_{\bm p}
    S^{(s)}_{ab}(\bm p,\omega)
    (\bm\alpha\cdot\bm p)^2
    \frac{\partial^2}{\partial\mu_\tau^2}
    \Theta\!\left(\mu_\tau-E_{p\tau}\right)
   \nonumber\\&=\delta_{ab}\Lambda^{(s)}_{0}(\omega) +
    \frac{\alpha^2}{2}
    \frac{\partial}{\partial\mu_\tau}
    \int_{\bm p} p^2
    S^{(s)}_{ab}(\bm p,\omega)
    \cos^2(\phi-\vartheta)
    \delta\!\left(\mu_\tau-E_{p\tau}\right).
\end{align}
Writing \(\bm\alpha=\alpha(\cos\vartheta,\sin\vartheta)\), the angular
integration gives the anisotropic tensor structure
\begin{align}\label{sm_eq:Lambda_inter}
  \Lambda^{(s)}_{ab}(\omega)
  = \delta_{ab}\left[\Lambda^{(s)}_{0}(\omega) + \frac{\alpha^2}{2}{\cal C}^{(s)}_{0}(\omega) \right]+ \frac{\alpha^2}{2} {\cal C}^{(s)}(\omega)
  \begin{pmatrix}
       \cos2\vartheta &   \sin 2\vartheta
      \\
       \sin2\vartheta & -\cos2\vartheta
  \end{pmatrix}_{ab},
\end{align}
where 
\begin{align}
    {\cal C}^{(s)} (\omega)
    &=\sum_\tau
    \frac{1}{(2\pi\hbar)^2}
    \frac{\partial}{\partial\mu_\tau}
    \int^{2\pi}_0 d\phi \int_0^\infty dp\,
    p^3 S^{(s,\tau)}(p,\phi,\omega) 
    \delta\!\left(\mu_\tau-E_{p\tau}\right)
    \nonumber\\
    &=\sum_\tau
    \frac{1}{(2\pi\hbar)^2 v^4}
     \frac{\partial}{\partial\mu_\tau} \int^{2\pi}_0 d\phi
    \int_{|m_\tau|}^{\infty} d\epsilon\,
    \epsilon(\epsilon^2-m_\tau^2)
    S^{(s,\tau)}(\sqrt{\epsilon^2-m^2_\tau}/v,\phi,\omega)
    \delta\!\left(\mu_\tau-\epsilon\right)
    \nonumber\\
    &=\sum_\tau
    \frac{\Theta(\mu_\tau-|m_\tau|)}{(2\pi\hbar)^2 v^4}
    \frac{\partial}{\partial\mu_\tau}
    \left[
    \mu_\tau(\mu_\tau^2-m_\tau^2) \int^{2\pi}_0 d\phi\,
    S^{(s,\tau)}(\sqrt{\mu^2_\tau-m^2_\tau}/v,\phi,\omega)
    \right].
\end{align}
Here \(S^{(s,\tau)}(p,\phi,\omega)\) denotes the part of the untilted
interband kernel which contributes to the traceless anisotropic projection of
\(S^{(s)}_{ab}\) at each valley. 
Defining
\begin{align}
x_\tau=\frac{\Omega}{\mu_\tau},
\qquad
y_\tau=\frac{m_\tau}{\mu_\tau},
\qquad
\Omega=\hbar\omega+i\gamma ,
\end{align}
and after a lengthy but straightforward calculation, we obtain 
\begin{align}
{\cal C}^{(s)}(\omega)
=
-\sum_\tau \frac{i \tau e^3}{32\pi\mu_\tau^{2}}
\Theta(1-|y_\tau|) \frac{1-y_\tau^{2}}{\left(x_\tau^{2}-4\right)^{4}}
\left[
x_\tau\,f(x_\tau,y_\tau)
+
8s y_\tau\,g(x_\tau,y_\tau)
\right],
\end{align}
where 
\begin{align}
&f(x,y)
=
-960
-160y^{2}
+8960y^{4}
+\left(
-160
+816y^{2}
-3920y^{4}
\right)x^{2}
+\left(
4
-140y^{2}
+696y^{4}
\right)x^{4}
+\left(
9y^{2}
-45y^{4}
\right)x^{6},
\\
&g(x,y)
=
480
-1440y^{2}
+\left(80+320y^{2}\right)x^{2}
-\left(2+50y^{2}\right)x^{4}
+3y^{2}x^{6}.
\end{align}
After including the spin-degeneracy factor $g_s=2$, Eq.~\eqref{sm_eq:Lambda_inter} yields the tilt dependence in Eq.~(24) and the interband contributions to Eqs.~(25) and (26) of the main text.

\subsection{Boltzmann analysis of the Fermi-surface contribution}

We now derive the intraband Fermi-surface contribution to the nonlocal
elasto-optic coefficient. In the semiclassical limit, the strain-induced
pseudo-gauge field enters through the local band energy and velocity,
\begin{align}
    \tilde\varepsilon_{\lambda\tau}(\bm p,\bm R)
    &=
    \varepsilon_{\lambda\tau}(\bm p)
    +
    {\cal A}_a(\bm R)g^a_{\lambda\tau}(\bm p),
    \label{eq:S_boltz_energy_shift}
    \\
    \tilde v^{\lambda\tau}_b(\bm p,\bm R)
    &=
    v^{\lambda\tau}_b(\bm p)
    +
    {\cal A}_a(\bm R)w^{a,\lambda\tau}_b(\bm p).
    \label{eq:S_boltz_velocity_shift}
\end{align}
Here
\begin{equation}
    \varepsilon_{\lambda\tau}(\bm p)
    =
    \tau\bm\alpha\cdot\bm p
    +
    \lambda E_{\tau p},
    \qquad
    E_{\tau p}=\sqrt{v^2p^2+m_\tau^2},
    \label{eq:S_boltz_band_energy}
\end{equation}
where \(\lambda=\pm\) labels the conduction and valence bands and
\(\tau=\pm\) labels the valleys. Since \(\bm p\) denotes physical momentum,
\begin{equation}
    v^{\lambda\tau}_b
    =
    \partial_{p_b}\varepsilon_{\lambda\tau},
    \qquad
    w^{a,\lambda\tau}_b
    =
    \partial_{p_b}g^a_{\lambda\tau}.
    \label{eq:S_boltz_v_w_def}
\end{equation}
For the pseudo-gauge coupling \(-ev\tau\bm\sigma_\tau\), the intraband
projected vertex is
\begin{equation}
    g^a_{\lambda\tau}(\bm p)
    =
    \langle u_{\lambda\tau\bm p}|
    (-ev\tau\sigma^a_\tau)
    |u_{\lambda\tau\bm p}\rangle
    =
    -\lambda\tau e v^2\frac{p_a}{E_{\tau p}} .
    \label{eq:S_boltz_g_vertex}
\end{equation}
In the following derivation, band, valley, and momentum labels are suppressed
where no confusion can arise.

A massive Dirac band also carries Berry curvature and therefore supports a
local anomalous Hall response. This local contribution can generate
SAM-sensitive terms, but it does not contribute to the leading
\(Q_b{\cal A}_a(\bm k)\) tensor responsible for the OAM-odd response considered
here. The optical anomalous velocity produces a local
\(E_j{\cal A}_a\) response rather than an optical-spatial-dispersion term. We
therefore keep only the ordinary band velocity in the intraband calculation.
The distribution function
\begin{equation}
      f=f_0(\xi)+\delta f,
    \qquad
    \xi=\varepsilon-\mu,  
\end{equation}
obeys the relaxation-time Boltzmann equation
\begin{equation}
    \partial_t f
    +
    \tilde{\bm v}\cdot\nabla_{\bm R}f
    +
    \dot{\bm p}\cdot\nabla_{\bm p}f
    =
    -\frac{\gamma}{\hbar}\,\delta f ,
    \label{eq:S_boltz_equation}
\end{equation}
where
\begin{equation}
    f_0(\xi)
    =
    \frac{1}{1+e^{\xi/T}}
    \label{eq:S_boltz_fermi_function}
\end{equation}
is the equilibrium Fermi--Dirac distribution. Here \(\gamma\) is the momentum
relaxation broadening and has dimensions of energy. We use the retarded
energy variable
\begin{equation}
    \Omega=\hbar\omega+i\gamma,
    \qquad
    D_\omega=\frac{i}{\Omega}.
    \label{eq:S_boltz_Omega_D_def}
\end{equation}
Equivalently,
\begin{equation}
     D_\omega
    =
    \frac{1}{\gamma-i\hbar\omega}.
\end{equation}
The force contains both the optical electric drive and the strain-gradient
force,
\begin{equation}
    \dot{\bm p}
    =
    -e\bm E(\bm R,t)
    -
    \nabla_{\bm R}
    \left[
        {\cal A}_a(\bm R)g^a(\bm p)
    \right].
    \label{eq:S_boltz_force}
\end{equation}
The second term in Eq.~\eqref{eq:S_boltz_force} is proportional to
\(k_b{\cal A}_a(\bm k)\) in momentum space and belongs to the strain-gradient
channel. Since the OAM response discussed in the main text is the
optical-spatial-dispersion channel proportional to
\(Q_b{\cal A}_a(\bm k)\), we evaluate the material kernel at \(k=0\) and omit
this explicit force term below.

We solve Eq.~\eqref{eq:S_boltz_equation} perturbatively in the optical field
and in the static strain texture. In the absence of strain, the correction
linear in the optical field satisfies the usual drift equation. For an optical
Fourier component with momentum \(\bm Q\) and frequency \(\omega\), one obtains
\begin{equation}
    \delta f^{E}
    =
    eE_j
    \frac{
        \hbar v_j f_0'
    }{
        \gamma-i\hbar\omega+i\hbar\bm Q\cdot\bm v
    },
    \label{eq:S_boltz_delta_f_E}
\end{equation}
where \(f_0'\equiv\partial f_0/\partial\xi\). The finite optical momentum in
the denominator of Eq.~\eqref{eq:S_boltz_delta_f_E} is the origin of the
spatially dispersive response. 
We expand the denominator for small $\mathbf {Q}$ which gives 
\begin{align}
    \left(
        \gamma-i\hbar\omega+i\hbar\bm Q\cdot\bm v
    \right)^{-1}
    &=
    \frac{i}{\Omega-\hbar\bm Q\cdot\bm v}
    =
    \frac{i}{\Omega}
    +
    \frac{i\hbar Q_bv_b}{\Omega^2}
    +
    O(Q^2).
    \label{eq:S_boltz_denominator_expansion}
\end{align}
The \(Q\)-linear optical correction is therefore
\begin{equation}
    \delta f^{QE}
    =
    i e\hbar^2 Q_bE_j
    \frac{v_bv_j}{\Omega^2}
    f_0'
    =
    -ie\hbar^2 Q_bE_jD_\omega^2v_bv_jf_0' .
    \label{eq:S_boltz_delta_f_QE}
\end{equation}

We next turn on the strain field and keep only terms linear in both
\(E_j\) and \({\cal A}_a\). Since the material kernel is evaluated in the
long-wavelength-strain limit, the strain momentum is set to zero inside the
kernel, while the external amplitude \({\cal A}_a(\bm k)\) is retained. The
mixed correction obeys
\begin{align}
    \left(
        \gamma-i\hbar\omega+i\hbar\bm Q\cdot\bm v
    \right)
    \delta f^{E{\cal A}}
    &=
    \hbar eE_j{\cal A}_a
    \left(
        v_jg^a f_0''
        +
        w_j^a f_0'
    \right)
    -
    i\hbar{\cal A}_a Q_b w_b^a\,\delta f^E .
    \label{eq:S_boltz_delta_f_EA_eq}
\end{align}
The two terms inside the first parenthesis in
Eq.~\eqref{eq:S_boltz_delta_f_EA_eq} come from the strain-induced shift of the
band energy and from the strain correction to the band velocity, respectively.
The last term comes from the strain correction to the streaming operator,
\({\cal A}_a\bm w^a\cdot\nabla_{\bm R}\delta f^E\). Terms generated by the
explicit strain-gradient force in Eq.~\eqref{eq:S_boltz_force} are proportional
to \(k_b{\cal A}_a(\bm k)\) and are not included in this
\(Q_b{\cal A}_a\) channel.
Keeping only the part proportional to \(Q_bE_j{\cal A}_a\), one obtains
\begin{equation}
    \delta f^{QE{\cal A}}
    =
    i e\hbar^2 Q_bE_j{\cal A}_a
    \frac{1}{\Omega^2}
    \left[
        \left(
            v_bw_j^a+w_b^av_j
        \right)f_0'
        +
        v_jv_bg^af_0''
    \right],
    \label{eq:S_boltz_delta_f_QEA_Omega}
\end{equation}
or, equivalently,
\begin{equation}
    \delta f^{QE{\cal A}}
    =
    -ie\hbar^2 Q_bE_j{\cal A}_aD_\omega^2
    \left[
        \left(
            v_bw_j^a+w_b^av_j
        \right)f_0'
        +
        v_jv_bg^af_0''
    \right].
    \label{eq:S_boltz_delta_f_QEA}
\end{equation}
This is the only mixed distribution correction needed for the leading
OAM-sensitive intraband response.
The current must also be expanded to first order in the strain correction to
the velocity. To order \(Q_bE_j{\cal A}_a\),
\begin{equation}
    j_i(\bm Q,\omega)
    =
    -e
    \sum_{\lambda,\tau}
    \int_{\bm p}
    \left[
        v_i\,\delta f^{QE{\cal A}}
        +
        {\cal A}_a w_i^a\,\delta f^{QE}
    \right],
    \label{eq:S_boltz_current_before_parts}
\end{equation}
where
\begin{equation}
    \int_{\bm p}
    \equiv
    \int\frac{d^2p}{(2\pi\hbar)^2}.
\end{equation}
The first term in Eq.~\eqref{eq:S_boltz_current_before_parts} is the ordinary
velocity acting on the mixed distribution, while the second is the
strain-induced velocity acting on the spatially dispersive optical
distribution.
The term proportional to \(f_0''\) is not an independent Fermi-sea
contribution. Using \(\partial_{p_b}\xi_{\lambda\tau}=v_b\), it can be reduced
by integration by parts:
\begin{equation}
    \int_{\bm p}v_iv_jv_bg^af_0''
    =
    \int_{\bm p}v_iv_jg^a\,\partial_{p_b}f_0'
    =
    -
    \int_{\bm p}
    f_0'\,
    \partial_{p_b}
    \left(
        v_iv_jg^a
    \right).
    \label{eq:S_boltz_parts_identity}
\end{equation}
Since \(w_b^a=\partial_{p_b}g^a\), the term generated by differentiating
\(g^a\) in Eq.~\eqref{eq:S_boltz_parts_identity} cancels the explicit
\(v_iv_jw_b^af_0'\) contribution in the current. The result is a pure
Fermi-surface expression,
\begin{equation}
    j_i(\bm Q,\omega)
    =
    \sum_{\lambda,\tau}
    \int_{\bm p}
    f_0'(\xi_{\lambda\tau})
    K_{ijab}(\bm p,\omega)\,
    E_j{\cal A}_a Q_b ,
    \label{eq:S_boltz_current_kernel}
\end{equation}
with
\begin{align}
    K_{ijab}(\bm p,\omega)
    =
    ie^2\hbar^2D_\omega^2
    \bigg[
        &
        v_b
        \left(
            v_i\partial_{p_j}g^a
            +
            v_j\partial_{p_i}g^a
        \right)
        -
        g^a
        \left(
            v_i\partial_{p_b}v_j
            +
            v_j\partial_{p_b}v_i
        \right)
    \bigg].
    \label{eq:S_boltz_K_ijab}
\end{align}
Equivalently, in \(\Omega\) notation,
\begin{align}
    K_{ijab}(\bm p,\omega)
    =
    -\frac{ie^2\hbar^2}{\Omega^2}
    \bigg[
        &
        v_b
        \left(
            v_i\partial_{p_j}g^a
            +
            v_j\partial_{p_i}g^a
        \right)
        -
        g^a
        \left(
            v_i\partial_{p_b}v_j
            +
            v_j\partial_{p_b}v_i
        \right)
    \bigg].
    \label{eq:S_boltz_K_ijab_Omega}
\end{align}
Comparing Eq.~\eqref{eq:S_boltz_current_kernel} with
\(j_i \sim Q_bE_j{\cal A}_a\zeta_{ijab}\), we obtain
\begin{equation}
    \zeta_{ijab}(\omega)
    =
    \sum_{\lambda,\tau}
    \int_{\bm p}
    f_0'(\xi_{\lambda\tau})
    K_{ijab}(\bm p,\omega).
    \label{eq:S_boltz_zeta_final}
\end{equation}
The kernel in Eq.~\eqref{eq:S_boltz_K_ijab} is symmetric in the optical
indices \(i,j\). Therefore the circular-polarization projection
\(\Lambda^{(s)}_{ab}=e^*_{s,i}\zeta_{ijab}e_{s,j}\) has no helicity-odd part.
The intraband Boltzmann contribution is independent of the SAM index \(s\);
the contrast generated by this channel is therefore purely OAM-odd. We hence
write
\begin{equation}
    \Lambda_{ab}(\omega)
    =
    \frac{1}{2}
    \sum_{\lambda,\tau}
    \int_{\bm p}
    f_0'(\xi_{\lambda\tau})
    \left[
        K_{xxab}(\bm p,\omega)
        +
        K_{yyab}(\bm p,\omega)
    \right].
    \label{eq:S_boltz_B_projected}
\end{equation}
At zero temperature, we have
\begin{equation}
    f_0'(\xi_{\lambda\tau})
    =
    -\delta(\xi_{\lambda\tau}),
    \qquad
    \xi_{\lambda\tau}
    =
    \lambda E_{\tau p}
    +
    \tau\alpha_a p_a
    -
    \mu_\tau .
    \label{eq:S_boltz_zero_T_fprime}
\end{equation}
Expanding the Fermi-surface factor to first order in the tilt gives
\begin{align}
    f_0'(\xi_{\lambda\tau})
    &=
    -\delta
    \left(
        \lambda E_{\tau p}-\mu_\tau+\tau \bm\alpha \cdot \bm p
    \right)
    \approx
    -\delta(\lambda E_{\tau p}-\mu_\tau)
    -
    \tau\bm\alpha\cdot\bm p\,
    \delta'(\lambda E_{\tau p}-\mu_\tau)
    +
    O(\alpha^2).
    \label{eq:S_boltz_fprime_tilt_expansion}
\end{align}
Writing the tilt vector as
\begin{equation}
    \bm\alpha
    =
    \alpha(\cos\vartheta,\sin\vartheta),
    \label{eq:S_boltz_tilt_vector}
\end{equation}
we have
\begin{equation}
    \bm\alpha \cdot \bm p
    =
    \alpha p\cos(\phi-\vartheta)
    =
    \frac{\alpha}{v}
    \sqrt{E_{\tau p}^2-m_\tau^2}
    \cos(\phi-\vartheta).
    \label{eq:S_boltz_tilt_projection}
\end{equation}
Thus
\begin{align}
    f_0'(\xi_{\lambda\tau})
    \approx
    -\delta(\lambda E_{\tau p}-\mu_\tau)
    -
    \frac{\tau\alpha}{v}
    \sqrt{E_{\tau p}^2-m_\tau^2}
    \cos(\phi-\vartheta)
    \delta'(\lambda E_{\tau p}-\mu_\tau)
    +
    O(\alpha^2).
    \label{eq:S_boltz_fprime_tilt_expansion_polar}
\end{align}
We next expand the projected kernel in powers of the tilt,
\begin{align}
    \frac{
        K_{xxab}(\bm p,\omega)
        +
        K_{yyab}(\bm p,\omega)
    }{2}
    &=
    \alpha S^{(1)}_{\lambda\tau,ab}(\bm p,\omega)
    +
    \alpha^2 S^{(2)}_{\lambda\tau,ab}(\bm p,\omega)
    +
    O(\alpha^3).
    \label{eq:S_boltz_S_expansion}
\end{align}
The untilted contribution does not enter the anisotropic OAM channel. Keeping
the terms of order \(\alpha^2\), Eqs.~\eqref{eq:S_boltz_B_projected} and
\eqref{eq:S_boltz_fprime_tilt_expansion_polar} give 
\begin{align}
    \Lambda^{(s)}_{ab}
    =
    -\alpha^2
    \sum_{\lambda,\tau}
    \int_{\bm p}
    S^{(1)}_{\lambda\tau,ab}(\bm p,\omega)
    \frac{\tau}{v}
    \sqrt{E_{\tau p}^2-m_\tau^2}
    \cos(\phi-\vartheta)
    \delta'(\lambda E_{\tau p}-\mu_\tau)
    -
    \alpha^2
    \sum_{\lambda,\tau}
    \int_{\bm p}
    S^{(2)}_{\lambda\tau,ab}(\bm p,\omega)
    \delta(\lambda E_{\tau p}-\mu_\tau).
    \label{eq:S_boltz_B_u2_before_FS}
\end{align}
For electron doping, \(\mu_\tau>|m_\tau|\), only the conduction band
\(\lambda=+1\) contributes. We define
\begin{equation}
    p_{F,\tau}
    =
    \frac{\sqrt{\mu_\tau^2-m_\tau^2}}{v},
    \qquad
    \mu_\tau>|m_\tau|.
    \label{eq:S_boltz_pF_def}
\end{equation}
The radial delta function is reduced using
\begin{equation}
    \delta(E_{\tau p}-\mu_\tau)
    =
    \frac{\mu_\tau}{v^2p_{F,\tau}}
    \delta(p-p_{F,\tau}),
    \qquad
    E_{\tau p}=\sqrt{v^2p^2+m_\tau^2}.
    \label{eq:S_boltz_delta_radial}
\end{equation}
Hence, for any smooth function \(F_\tau(p,\phi)\),
\begin{align}
    \int_{\bm p}
    F_\tau(p,\phi)
    \delta(E_{\tau p}-\mu_\tau)
    =
    \int_0^{2\pi}
    \frac{d\phi}{(2\pi\hbar)^2}
    \frac{\mu_\tau}{v^2}
    F_\tau(p_{F,\tau},\phi).
    \label{eq:S_boltz_I0_functional}
\end{align}
Similarly, since
\begin{equation}
    \delta'(E_{\tau p}-\mu_\tau)
    =
    -
    \frac{d}{d\mu_\tau}
    \delta(E_{\tau p}-\mu_\tau),
    \label{eq:S_boltz_delta_prime_identity}
\end{equation}
we obtain
\begin{align}
    \int_{\bm p}
    F_\tau(p,\phi)
    \delta'(E_{\tau p}-\mu_\tau)
    =
    -
    \frac{d}{d\mu_\tau}
    \left[
        \int_0^{2\pi}
        \frac{d\phi}{(2\pi\hbar)^2}
        \frac{\mu_\tau}{v^2}
        F_\tau(p_{F,\tau},\phi)
    \right].
    \label{eq:S_boltz_I1_functional}
\end{align}
Applying Eqs.~\eqref{eq:S_boltz_I0_functional} and
\eqref{eq:S_boltz_I1_functional} to
Eq.~\eqref{eq:S_boltz_B_u2_before_FS}, with
\begin{equation}
       F_\tau^{(1)}(p,\phi)
    =
    \tau
    S^{(1)}_{\tau,ab}(p,\phi,\omega)
    p\cos(\phi-\vartheta), 
\end{equation}
gives the compact Fermi-surface expression
\begin{align}
    \Lambda_{ab}^{(s)}
    &=
    \alpha^2
    \sum_\tau
    \tau
    \frac{d}{d\mu_\tau}
    \left[
        \int_0^{2\pi}
        \frac{d\phi}{(2\pi\hbar)^2}
        \frac{\mu_\tau p_{F,\tau}}{v^2}
        S^{(1)}_{\tau,ab}(p_{F,\tau},\phi,\omega)
        \cos(\phi-\vartheta)
    \right]
    \nonumber\\
    &\quad
    -
    \alpha^2
    \sum_\tau
    \int_0^{2\pi}
    \frac{d\phi}{(2\pi\hbar)^2}
    \frac{\mu_\tau}{v^2}
    S^{(2)}_{\tau,ab}(p_{F,\tau},\phi,\omega).
    \label{eq:S_boltz_B_u2_FS_final}
\end{align}
%
The radial Jacobian gives the factors
\(\mu_\tau/v^2\) and \(\mu_\tau p_{F,\tau}/v^2\). The first term in
Eq.~\eqref{eq:S_boltz_B_u2_FS_final} is the Fermi-surface-shift contribution
from the tilt dependence of \(f_0'\), while the second is the explicit
\(\alpha^2\) contribution from the projected kernel.
Carrying out the remaining angular integrals gives
\begin{align}
    \Lambda^{\rm intra}_{ab}(\omega)
    =
    -\frac{i e^3 \alpha^2}{8\pi\Omega^2}
    \left[
        F(y_+)-F(y_-)
    \right]\delta_{ab}
    -
    \frac{i e^3 \alpha^2}{8\pi\Omega^2}
    \left[
        F(y_+)-F(y_-)
    \right]
    \begin{pmatrix}
        \cos2\vartheta &  \sin2\vartheta \\[4pt]
        \sin2\vartheta & -\cos2\vartheta
    \end{pmatrix}_{ab},
    \label{eq:S_boltz_B_final}
\end{align}
where
\begin{equation}
    y_\tau=\frac{m_\tau}{\mu_\tau},
    \label{eq:S_boltz_z_Omega_def}
\end{equation}
and
\begin{equation}
    F(y)
    =
    \Theta(1-|y|)
    \left(1-y^2\right)
    \left(1+3y^2\right).
    \label{eq:S_boltz_F_def}
\end{equation}
After including the spin-degeneracy factor $g_s=2$, Eq.~\eqref{eq:S_boltz_B_final} yields the intraband contribution to Eq.~(25) of the main text.